\documentclass{article}
\usepackage[labelfont=bf]{caption} %
\usepackage[font={small}]{caption}   %

\usepackage{graphicx}%
\usepackage{dcolumn}%
\usepackage{bm}%
\usepackage[pdftex,colorlinks=true,linkcolor=blue,citecolor=blue]{hyperref}
\usepackage[utf8]{inputenc}
\usepackage{lipsum}  
\usepackage{amsmath}
\usepackage{authblk}
\usepackage{amsfonts}
\usepackage{placeins}
\usepackage[numbers, sort&compress]{natbib}  %
\usepackage{bibunits}
\usepackage{amssymb}
\usepackage{pdflscape} %

    \usepackage[table, usenames, dvipsnames]{xcolor}
    \usepackage{makecell}
    \usepackage{boldline}
    \setcellgapes{3pt}
\usepackage{amsmath}
\usepackage{amssymb}

\usepackage{textgreek}      %

\usepackage[strict]{changepage} %
\usepackage{nicefrac}

\defaultbibliographystyle{naturemag}
\defaultbibliography{thesisrefsol}

    \usepackage{soul}   %
    \setstcolor{red} %
    \setul{}{1.5pt}  %

\usepackage{lineno}

\usepackage{xspace}
\newcommand{\TEzz}{TE\ensuremath{_\textrm{00}}\xspace}
\newcommand{\TMzz}{TM\ensuremath{_\textrm{00}}\xspace}

\title{Electro-optic non-reciprocal polarization rotation in lithium niobate}

\author{
    Oğulcan E. Örsel $^1$, Gaurav Bahl $^2$ \\
    $^1$ Department of Electrical $\&$ Computer Engineering, \\
    $^2$ Department of Mechanical Science $\&$ Engineering, \\
    University of Illinois at Urbana–Champaign, Urbana, IL 61801 USA \\
}

\date{}

\begin{document}
\begin{bibunit}

\maketitle

\begin{abstract}

Polarization is a fundamental degree of freedom for light and is widely leveraged in free space and fiber optics. Non-reciprocal polarization rotation, enabled via the magneto-optic Faraday effect, has been essentially unbeatable for broadband isolators and circulators. For integrated photonics foundries, however, there is still no good path to producing low-loss magneto-optic components, which has prompted a search for alternatives that do not use polarization rotation. Moreover, magneto-optic materials tend to be highly lossy, and while large ($10-100$ rad/cm) polarization rotation can be achieved, the key figure of merit (rotation-per-loss) is typically $< 1$ rad/dB. Here, we demonstrate that broadband non-reciprocal polarization rotation can be produced using electro-optics in nanophotonic devices. Our demonstration leverages electro-optic inter-polarization scattering around 780 nm in lithium niobate, in which the reciprocity is broken with the help of a radiofrequency stimulus that carries synthetic momentum. While the demonstrated electro-optic polarization rotation rate is $\approx 1$ rad/cm, the exceptionally low loss of lithium niobate enables non-reciprocal polarization rotators with figures of merit that are 1-2 orders-of-magnitude better than what is possible with magneto-optics. This approach can be replicated in III-V platforms paving the way for high-performance lasers with co-integrated monolithic isolators.

\end{abstract}

\maketitle

\vspace{12pt}

The Magneto-optic Faraday rotation effect (MOFE) is the foundation for all commercially-fielded isolator and circulator devices available today. 
Its primary action is to induce directional non-reciprocity in the polarization degree of freedom for photons~\cite{Bennett_Stern_1965,Hulme_Fowler_1932,Serber_1932}, which when combined with polarization filters can produce a direction dependent power-transmission function~\cite{Aplet_64,S_Fischer_1987,Wolfe_Hegarty_Dillon_Luther_Celler_Trimble_Dorsey_1985}.
The key feature that enables the dominance of the MOFE in non-reciprocal technologies is the very wide operating bandwidth over which it acts, and the strength of the polarization rotation (expressed as the Verdet constant in rad/T$\cdot$cm) in magneto-optic materials, e.g. terbium gallium garnet (TGG) being a prominent example. 
For these reasons, MOFE-based circulators are found in all complex optical systems, and isolators are considered essential for laser stabilization.

MOFE-based nonreciprocal devices do come with limitations, however, as they require specialized materials and magnetic biasing, neither of which are typically available in integrated photonics platforms. 
The effect also tends to be strongly chromatic, which forces changes to material selection and adjustment of the magnetic bias in order to tune the operational wavelength.
Finally, magneto-optic materials also tend to be quite lossy with typical propagation loss approaching 50-70 dB/cm ~\cite{Zhang:19,Yan:20}. This compels designers to only use the minimum amount of material and requires careful management between the strength of the polarization non-reciprocity generated (rad/cm) and the signal attenuation incurred (dB/cm), with the central figure of merit being their ratio (rad/dB). 
While a number of successful attempts have been made in producing on-chip MOFE isolators~\cite{Srinivasan_Stadler_2018,Ross:11,Ghosh:12aa,Huang:17,Zhang:2017wq,Du:2018wy,Zhang:19,Yan:20}, the common theme is to use the minimum allowable magneto-optic material to minimize the accompanying losses.

Because of these limitations, multiple alternative approaches have been explored that would improve foundry compatibility, that rely on lithographical patterning for wavelength tuning rather than on materials changes, and that avoid magnetic fields for sensitive applications. 
	The most notable alternatives leverage spatio-temporal modulations or momentum biasing through optomechanical interactions ~\cite{Ruesink:16}, acousto-optics~\cite{Sohn18,Kittlaus_2018,Sohn:2019aa,Tian:2020ti,sarabalis2020,Kittlaus:2021wq,Sohn_Orsel_Bahl_2021}, and electro-optics~\cite{Sounas:14,Doerr:11,lira2012,Tzuang2014, Li:2014vo,Dostart:2021uh}.
A few of these approaches can produce near-ideal isolation behavior with simultaneously very low insertion loss and large contrast~\cite{Tian:2020ti,Sohn_Orsel_Bahl_2021}, and are therefore quite competitive with magneto-optics (for a recent detailed comparison we refer the reader to the Supplementary Information of Ref.~\cite{Sohn_Orsel_Bahl_2021}). 
Even so, the interaction that generates the non-reciprocity is usually weak, and resonant structures are often used to enhance the non-reciprocal effect but inadvertently limit the isolation bandwidth. Magneto-optics still remain unbeatable on the bandwidth metric.

In this work, we show that an MOFE-like nonreciprocal polarization rotation effect can be achieved on-chip using electro-optic materials, without the use of magneto-optics. 
Electro-optic materials such as lithium niobate (LiNbO\textsubscript{3}) have been rising to prominence due to their wide band gaps, extreme low loss~\cite{Desiatov:19,Sohn_Orsel_Bahl_2021}, and the possibility of producing active devices~\cite{Wang_2018,Zhang_comb_2019,Sohn_Orsel_Bahl_2021,Hu_Nat_2021,Zhu:21}. Indeed, these materials can be turned into dynamic polarization rotators via external perturbations~\cite{Campbell_Steier_1971,Noe_Smith_1988,Qin_Lu_Pollick_Sriram_Yoo_2017}. Unlike the MOFE, however, the physics of the electro-optic polarization rotation effect is reciprocal and cannot be used directly to produce non-reciprocal devices. 
We show here that this reciprocity issue is resolvable by introducing a large synthetic momentum bias into the electro-optic modulation.
As with previous approaches that use synthetic momenta \cite{lira2012,Sounas:14,Tian:2020ti}, here we demonstrate that this approach produces a very strong non-reciprocity in the polarization conversion.
Importantly, since high quality electro-optic materials can exhibit ultra-low propagation loss ($<$0.1 dB/cm) \cite{Desiatov:19,Sohn_Orsel_Bahl_2021}, we show that the figure of merit for electro-optic non-reciprocal polarization rotators can be 1-2 orders of magnitude greater than what is possible with the best MOFE-based on-chip devices to date~\cite{Zhang:19,Yan:20}.

\begin{figure}[htp]
    \begin{adjustwidth*}{-1in}{-1in}
    \hsize=\linewidth
    \includegraphics[width=1\columnwidth]{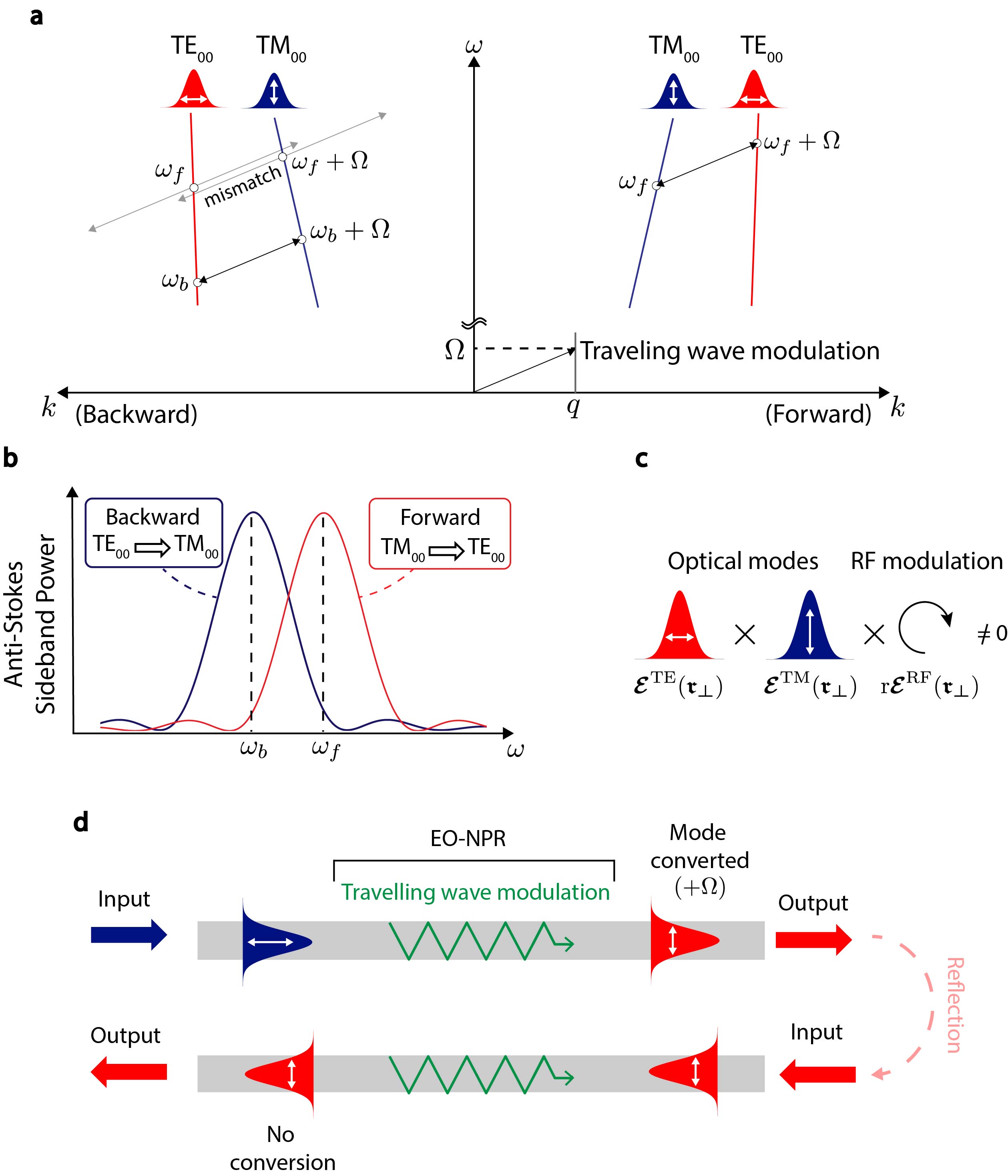}
    \centering
    \caption{
        \textbf{Principle of non-reciprocal polarization rotation via travelling wave induced interband scattering.}
        \textbf{(a)} We consider here the fundamental \TEzz and \TMzz modes of a waveguide with linear dispersion. A spatial modulation with large momentum applied to the waveguide has the ability to scatter photons between the modes and achieve polarization rotation. If this modulation additionally has a time-domain component, i.e. results in a traveling wave with frequency $\Omega$ and momentum $q$, the phase-matching is only satisfied for one direction of propagation at a selected optical probing frequency.
        \textbf{(b)} 
        The group velocities for the modes are not matched in the general case, which results in a sinc$^2$ spectral dependence of the scattering process, and a propagation direction-dependent spectral shift in the scattering profiles. Here we illustrate the anti-Stokes case only.
        \textbf{(c)} 
        The applied modulation must utilize suitable electro-optic coefficients $\textrm{r}$ that break the orthogonality of the polarizations.
        \textbf{(d)} One illustrative example of the resulting behavior of the electro-optic non-reciprocal polarization rotator (EO-NPR) is presented.
    }
    \label{fig:1}
    \end{adjustwidth*}
\end{figure}

\vspace{12pt}

The principle of our broadband nonreciprocal polarization rotator is described in Fig.~\ref{fig:1}.d
We consider an optical waveguide that supports two modes of types {\TEzz} and {\TMzz}, i.e. with mutually transverse polarization. These guided modes are orthogonal and have distinct dispersions that are separated in momentum-frequency space by a non-zero momentum. 
We can induce coupling between these modes by introducing a spatial perturbation that shifts the momentum for guided photons, such as a specialized grating, though the resulting polarization rotation is completely reciprocal \cite{Majumder_Shen_Polson_Menon_2017, Posner_Podoliak_Smith_Mennea_Horak_Gawith_Smith_Gates_2019}.
Non-reciprocal behavior can now be introduced by adding a time-dependent component to the perturbation, that is, by using a traveling wave modulation that shifts both frequency and momentum for light~\cite{Hwang:97,Yu_Fan_2009,Kim2015,Sohn18,Sohn_Orsel_Bahl_2021}.
With this type of perturbation applied, light injected into either the {\TMzz} or {\TEzz} mode can only be converted into the transverse polarization partner mode as long as the phase matching relations hold true, that is, both frequency and momentum matching conditions are satisfied (see Supplement \S\ref{sec:EOrate} $\&$ \S\ref{sec:Phasematching})
In the specific example shown in Fig.~{\ref{fig:1}}a, we apply a traveling wave modulation at frequency $\Omega$ with a forward-directed momentum such that it only permits conversion from TM polarization to TE for forward-propagating light via anti-Stokes scattering from input frequency $\omega_f$ to $\omega_f+\Omega$. Conversely, forward-propagating light in the TE polarization can only be converted to TM via Stokes scattering from $\omega_f+\Omega$ down to $\omega_f$. On the other hand, for backward-propagating light, the phase matching is satisfied differently and the frequency shift during scattering is reversed such that conversion from TM to TE polarization can only take place between frequencies $\omega_b + \Omega$ (for TM) and $\omega_b$ (for TE). 
As a general principle this asymmetric frequency shift effect is broadband, i.e. works for any $\omega_{f}$ and $\omega_{b}$, and extends over a frequency range where the group velocities of the optical modes are matched. 

In practice, however, these group velocities are rarely matched due to the dispersive nature of the material and the waveguide geometry. In order to understand the implications, we consider the anti-Stokes process in Fig.~{\ref{fig:1}}a although identical dynamics are produced for the Stokes process.
For light injected into the {\TMzz} mode in the forward direction, the anti-Stokes polarization conversion is maximized at $\omega_f$ where perfect phase matching occurs, and as a function of detuning the conversion exhibits a sinc$^2$ profile that originates due to phase accumulation (Fig.~\ref{fig:1}b). A detailed discussion is provided in the Supplement \S\ref{sec:Phasematching}. 
On the other hand, due to the mismatched dispersion, light entering the {\TEzz} mode in the backward direction can only optimally undergo anti-Stokes scattering back into the \TMzz mode at a different frequency $\omega_b$. This produces a direction-dependent spectral shift in the sinc$^2$ profile (Fig.~\ref{fig:1}b) and results in a nonreciprocal conversion process for the forward and backward directions. 
The spectral shift can be analytically evaluated as $\omega_{f}-\omega_{b} = \Omega \, (n_g^{\rm{TM}}+n_g^{\rm{TE}})/(n_g^{\rm{TM}}-n_g^{\rm{TE}})$, and increases with larger values of the applied modulation frequency $\Omega$ and with better matching of the group indices. In fact, if the group indices are exactly matched then the phase matching in the opposite direction is never satisfied and the bandwidth becomes infinite in theory with limits imposed by other practical constraints.
In the cases where this shift $\omega_{f}-\omega_{b}$ is well resolved, i.e. $(n_{g}^{\rm{TM}}+n_{g}^{\rm{TE}})\Omega \mathcal{L} \, / \, \pi c\gg1$ (see Supplement \S\ref{sec:Nonreciprocal} 
), where $\mathcal{L}$ is the total interaction length, the device can exhibit large mode conversion in the forward direction but negligible conversion in the backward direction over some frequency bands (Fig.~\ref{fig:1}d). 
Furthermore, this non-reciprocal behaviour is maximized if the spectral shift aligns the peak of the sinc$^2$ in one direction with a null in the opposite direction of propagation. 
The above analyses are discussed in detail in the Supplement \S\ref{sec:Nonreciprocal}. 

In addition to the above phase matching requirements, we must ensure a non-zero overlap integral between the transverse polarization modes and the traveling wave modulation that provides the rotation action. Such inter-polarization scattering can be achieved electro-optically \cite{Campbell_Steier_1971, Huang_Lin_Chen_Huang_2007, Ding_Zheng_Chen_2019}, i.e. using the second order optical nonlinearity which produces refractive index change under an applied radio-frequency (RF) electric field.
The overlap integral requirement can then be summarized as 
$\int \sum_{ijk} \mathcal{E}_i^\textrm{TM}(\boldsymbol{\mathfrak{r}_\bot}) \, \mathcal{E}_j^\textrm{TE}(\boldsymbol{\mathfrak{r}_\bot}) \,\textrm{r}_{ijk} \, \mathcal{E}_k^\textrm{RF}(\boldsymbol{\mathfrak{r}_\bot}) \, d\boldsymbol{\mathfrak{r}_\bot} \neq 0$
where $\boldsymbol{\mathcal{E}}^\textrm{TE}(\boldsymbol{\mathfrak{r}_\bot})$, $\boldsymbol{\mathcal{E}}^\textrm{TM}(\boldsymbol{\mathfrak{r}_\bot})$, and $\boldsymbol{\mathcal{E}}^\textrm{RF}(\boldsymbol{\mathfrak{r}_\bot})$ are the transverse field distribution of optical and applied RF modes respectively (see Supplement \S\ref{sec:EOrate}) and the subscripts ($ijk$) refer to coordinate axes for material's electro-optic tensor. 
This requirement is visually depicted in Fig.~\ref{fig:1}c.
The overlap integral can be interpreted into a specific requirement of non-zero electro-optic coefficients $\textrm{r}\textsubscript{ijk}$ for $i\neq j$~\cite{Weis:1985aa,Boyd}, which are significant in many electro-optic materials, such as LiNbO\textsubscript{3} \cite{Campbell_Steier_1971, Huang_Lin_Chen_Huang_2007, Ding_Zheng_Chen_2019}, GaAs \cite{McCallum_Huang_Smirl_Sun_Towe_1995,Suzuki_Tada_1984}, InP~\cite{Suzuki_Tada_1984} and InGaP~\cite{Ueno:97}.
These materials therefore give us a path to realize non-reciprocal polarization rotation in foundry compatible processes without requiring magneto-optics, by just applying appropriately designed RF stimuli. 
As an important observation, the best chip-scale magnetless optical isolators to date have used materials that are not easy to co-integrate with active III-V photonics, and this electro-optic approach could help traverse that barrier.

\begin{figure}[htp]
   \begin{adjustwidth*}{-1in}{-1in}
    \hsize=\linewidth
    \includegraphics[width=1.1\columnwidth]{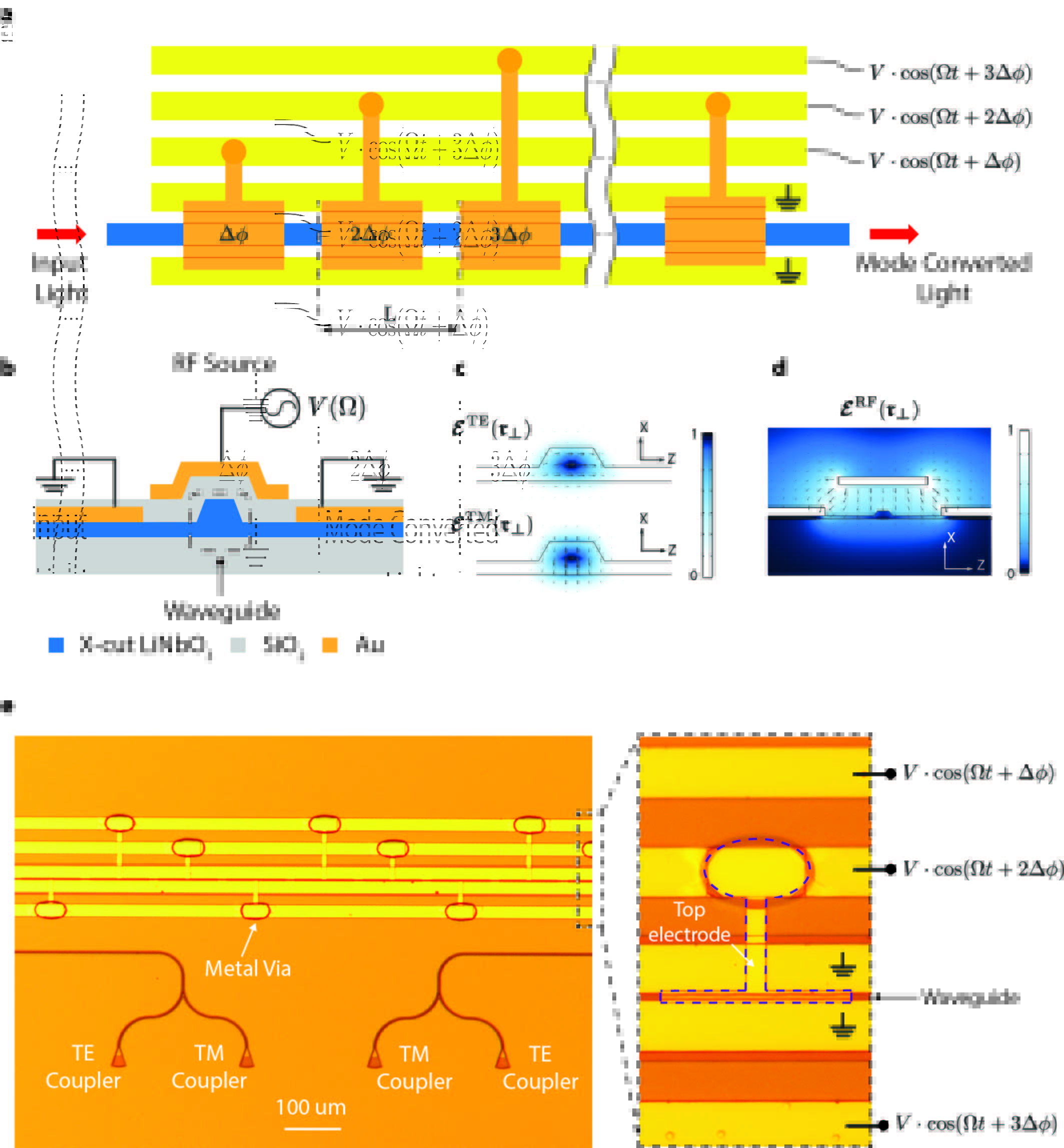}
    \centering
    \caption{
    \textbf{Implementation of a 780 nm electro-optic non-reciprocal polarization rotator (EO-NPR) in lithium niobate.}
    \textbf{(a)} A plan view of the EO-NPR device. The travelling wave modulation is realized by using a split electrode structure driven by three RF bus lines. The momentum is set by the relative phase shift $\Delta\phi$ irrespective of the frequency $\Omega$. The incoming photons within the waveguide (blue) are polarization-converted via electro-optic effect which is realized by the top (dark yellow) and bottom (light yellow) electrodes.  
    \textbf{(b)} Cross-sectional schematic of the device showing the electrodes (yellow), the lithium niobate (LN) waveguide (blue), and the oxide cladding (grey). The ground electrodes and the RF stimulus is also labeled to show the RF mode that is excited.
    \textbf{(c)} Finite element simulation of the \TEzz and \TMzz optical modes supported by the waveguide.
    \textbf{(d)} Simulated RF mode shows the vertically symmetric electric field profile that is required to realize polarization rotation.
    \textbf{(e)} True-color microscope image of the device. The LN nano-photonic structures appear in dark orange and the gold electrodes appear in yellow. The inset shows a zoomed-in view of a single electrode.
    }
    \label{fig:2}
    \end{adjustwidth*}
\end{figure}

In order to bridge the momentum gap between the optical modes, as described above, the spatiotemporally propagating RF field profile needs to exhibit a low group velocity, that is, large momentum for low frequency.
This is relatively simple if using acoustic phonons as in past studies \cite{Sohn18,Kittlaus_2018,Sohn:2019aa,sarabalis2020,Kittlaus:2021wq,Sohn_Orsel_Bahl_2021} as they can carry significant momentum.
Unfortunately, RF modes are not capable of providing sufficient momentum for modest frequencies~\cite{Ghione_2009,Wang_2018,Zhu:21} which sets a practical constraint to this approach.
To circumvent this issue we generate a synthetic momentum for the RF excitation along the waveguide by engineering a split electrode structure~\cite{lira2012,Sounas:14,Tian:2020ti,Kim:2021te}.
Each electrode (of pitch $L$) is provided a single tone RF stimulus at frequency $\Omega$ with a relative phase shift of $\Delta\phi$ between electrodes (see Fig.~\ref{fig:2}a). The dominant component of the resulting spatial phase profile produces an effective momentum at $q = \Delta\phi/L$ that can bridge the momentum gap and, in principle, could be tuned dynamically with a suitably adaptive design. A detailed discussion on this synthetic momentum, along with its primary and higher order Fourier components, is provided in the Supplement \S\ref{sec:Syntheticmomentum}. For this work, we employed a 3-phase periodicity to produce the required synthetic momentum.

\vspace{12pt}

For experimental demonstration we developed an electro-optic non-reciprocal polarization rotator (EO-NPR) near 780 nm, using a 300 nm X-cut LiNbO\textsubscript{3} on insulator (LNOI) integrated photonics platform.
Lithium niobate (LN) is an ideal material for these experiments since its wide bandgap provides a broad transparency window spanning 250 -- 5300 nm that enables integrated nanophotonics with low propagation loss ($<$0.1 dB/cm) \cite{Desiatov:19,Sohn_Orsel_Bahl_2021}, and its second-order optical non-linearities enable very efficient electro-optic modulation~\cite{Weis:1985aa}. 
A plan view of the EO-NPR device is presented in Fig.~{\ref{fig:2}}a and a cross-section of the waveguide presented in Fig.~\ref{fig:2}b. The waveguide was designed to support only {\TEzz} and {\TMzz} modes near 780 nm and details on the dimensions are provided in the Supplement \S\ref{sec:Dimensions}.
We oriented the waveguide along the Y-crystal direction (X-Z plane) to harness the r\textsubscript{42} coefficient of LN (the double index notation is explained in the Supplement \S\ref{sec:EOrate}).
At both ends of the waveguide we fabricated 50:50 Y-splitters followed by grating couplers that are customized to the TE and TM modes near 780 nm for selecting the polarization.
Electro-optic modulation is enabled through a stimulus electrode above the waveguide, with ground electrodes on either side of the waveguide.
Additional details on the fabrication are provided in Methods.
The optical modes supported by the waveguide are presented in Fig.~\ref{fig:2}c. We also optimized the electrode distance to mitigate any excessive optical loss while maintaining large electric field for electro-optic modulation (see Supplement \S\ref{sec:Dimensions}).
An RF voltage applied on the top electrode generates a symmetric vertical field profile shown in Fig.~\ref{fig:2}d which is required to activate r\textsubscript{42} coefficient of LiNbO\textsubscript{3}. This configuration also conveniently produces a mirror symmetric horizontal electric field which is important to suppress unwanted intramodal scattering due to the large r\textsubscript{33} coefficient of LiNbO\textsubscript{3}.
The top electrodes are driven using three bus lines that provide the input voltages $V \cdot \cos(\Omega t + \textrm{n}\Delta\phi)$ 
at a frequency of $\Omega$ with relative phase $\Delta\phi = 2\pi/3$ for n = 1, 2, 3 which ensures a periodic spatial phase.
Fig.~\ref{fig:2}e presents a microscope view of a portion of the device showing dedicated grating couplers used for selectively injecting TE and TM light and the electrode configuration used for generating the synthetic RF traveling wave. 
The electrode pitch was set to $L = 120$ um with a total length of the interaction region $\mathcal{L} = 1.08$ mm (3 modulation periods were implemented).

\begin{figure}[ht!]
    \begin{adjustwidth*}{-1in}{-1in}
    \hsize=\linewidth
    \includegraphics[width=\columnwidth]{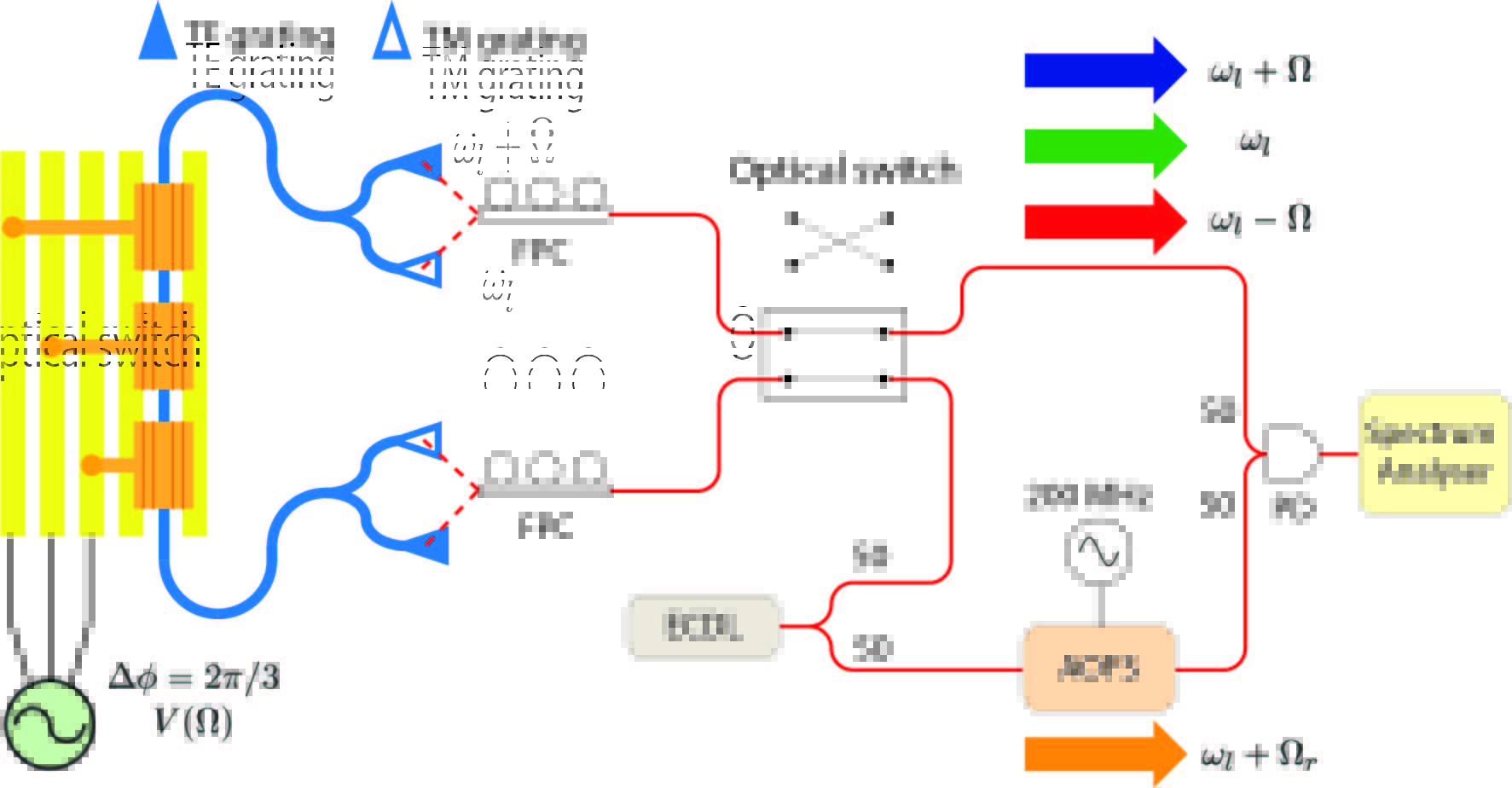}
    \centering
    \caption{
    \textbf{Experimental measurement setup.}
    Light (770-780 nm) from a fiber-coupled external cavity diode laser (ECDL) is split into a probe path and a reference path with 50:50 ratio. Fiber polarization controllers (FPCs) are used to adjust the light coupled into the chip prior to the grating couplers. An optical switch is used to change the measurement direction through the device. In the reference path, we introduce an acousto-optic frequency shifter (AOFS) to offset the frequency by 200 MHz, which enables heterodyne detection through a high-speed photodetector (PD) when the probe and reference paths are recombined. The on-chip electrodes are modulated using synchronized RF signal generators with a fixed relative phase offset. 
    }
    \label{fig:3}
    \end{adjustwidth*}
\end{figure}

We first characterize the primary radio-frequency (RF) and optical components of the EO-NPR.
The RF electrodes are individually characterised by a S-parameter measurement with a vector network analyzer.
The RF reflection spectrum (see Supplement \S\ref{sec:Electricalmodel}) shows an RC response with no resonances, which confirms that the excitation of each electrode is spatially uniform and non-propagating for the designed geometry.
This ensures that the split electrode structure will produce the desired momentum bias.
The typical insertion loss of the \TEzz and \TMzz gratings are identified by measuring the corresponding modal transmission through the waveguide with no stimulus applied, and we estimate the typical per-grating loss to be 6 dB and 11 dB for {\TEzz} and {\TMzz} modes respectively. Later we use these measurements in the Supplement \S\ref{sec:OpticalCharacterization} to calibrate the experimental results.

\begin{figure}[htp]
    \begin{adjustwidth*}{-1in}{-1in}
    \hsize=\linewidth
    \includegraphics[width=\columnwidth]{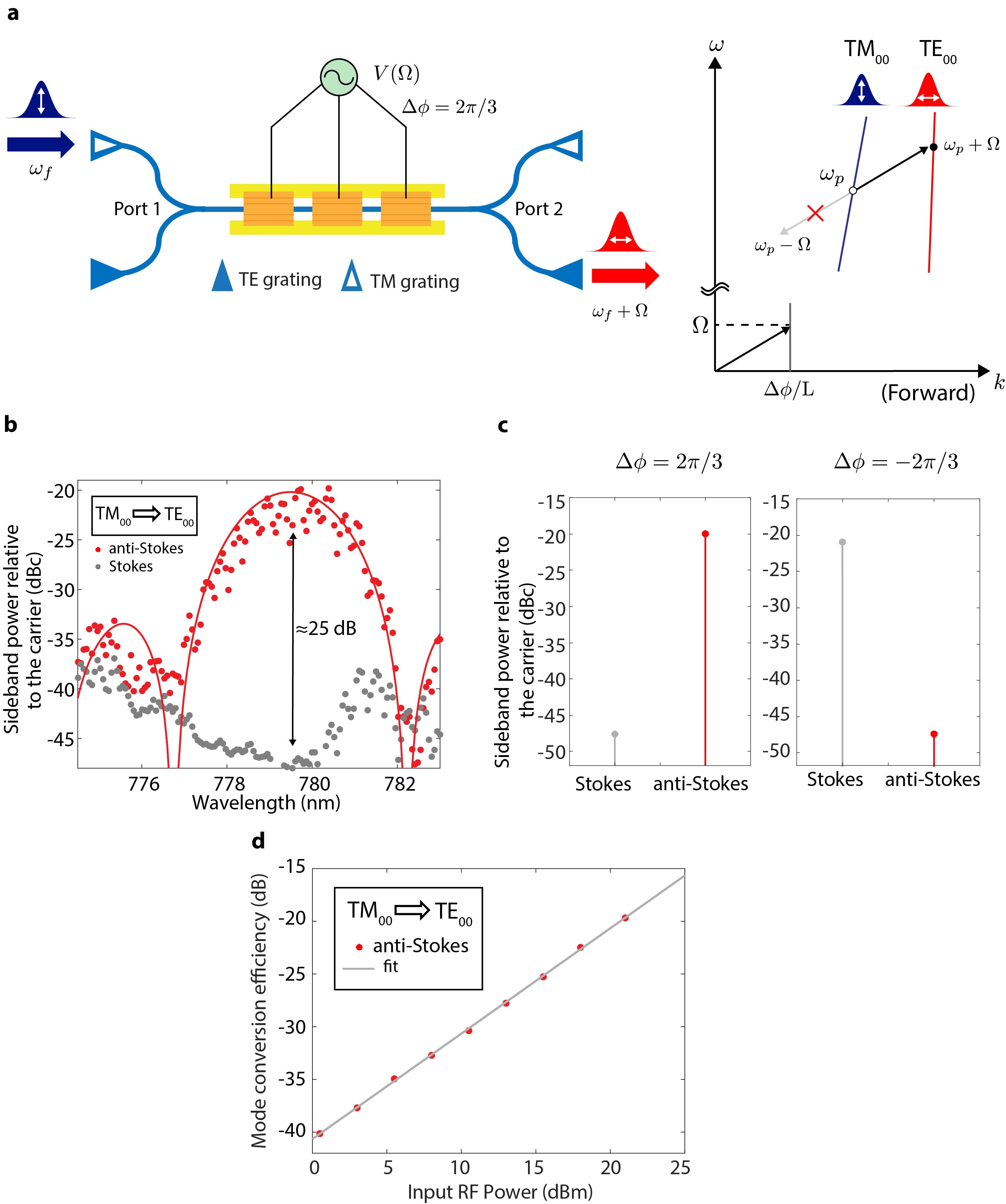}
    \centering
    \caption{
    \textbf{Experimental demonstration of single sideband inter-polarization scattering and broken time-reversal symmetry.} 
    \textbf{(a)} An RF traveling wave is applied to the EO-NPR via electrodes having 3-phase periodicity, which generates the required momentum ($\Delta\phi/L$). For $\Delta\phi=2\pi/3$ only anti-Stokes scattering is generated as illustrated in the band diagram since there are no available states for Stokes scattering.
    \textbf{(b)} With RF stimulus set at $\Omega = 2\pi\cdot3$ GHz, we measure the modulation efficiency into the \TEzz mode as a function of input wavelength. Since the phase matching condition changes with frequency, the modulation response displays a sinc$^2$ spectral characteristic (solid line is fit with theory, presented in Supplement \S\ref{sec:Phasematching}) for anti-Stokes scattering. Data normalization is discussed in the Supplement \S\ref{sec:OpticalCharacterization}.
    \textbf{(c)} The peak electro-optic inter-polarization scattering efficiency (occurs around 780 nm) measured for  applied $\Delta\phi = \pm 2\pi/3$ confirms the asymmetric scattering and the breaking of time-reversal symmetry. 
    \textbf{(d)} The modulation efficiency measured for different power levels shows good agreement with the theoretical prediction given by the finite element simulations. Here, the input RF power represents the total power divided into the three buses.
    }
    \label{fig:4}
    \end{adjustwidth*}
\end{figure}

In order to measure the electro-optic polarization rotation effect, we used an optical heterodyne measurement system (Fig.~\ref{fig:3}, details in Methods and \S\ref{sec:OpticalCharacterization}) which enables the separate characterization of input optical probes (i.e carrier) and their sidebands.
The polarization selective grating couplers additionally provide filtering action and assist with 
resolution of the inter-polarization scattering rate. 
We normalize all measurements with respect to the injected carrier power. 
The inter-polarization scattering experiment is depicted in Fig.~\ref{fig:4}a. An optical carrier signal $\omega\textsubscript{p}$ is injected into the waveguide {\TMzz} mode via port 1 while the output light is monitored on the \TEzz output at port 2.
A synthetically generated travelling RF wave is applied at 3 GHz at 21 dBm power level (each electrode) with a few different selections of relative phase $\Delta\phi$ as described below. The adjustment of this relative phase changes the momentum bias for a given frequency modulation, and alters the modulation process. Based on electrical calibrations in Supplementary \S\ref{sec:Electricalmodel} we estimate that this power level corresponds to $\approx$ 7.1 V applied on each electrode.

First, to observe the spectral dependence of this process which results from the group index difference between the modes, we measure the generated sidebands for different wavelengths (Fig.~\ref{fig:4}b).
For this demonstration, we selected the phase to be positive ($\Delta\phi = 2\pi/3$) so that the RF stimulus wave travels in the same direction as the optical carrier signal. %
This results in strong anti-Stokes scattering into the {\TEzz} mode since there is no density of states available for Stokes generation (Fig.~\ref{fig:4}a). 
In Fig.~\ref{fig:4}b, we see that the anti-Stokes signal is strongly produced around 780 nm and the modulation efficiency drops as we move away from this wavelength following the expected sinc$^2$ shape with a $\Delta\lambda\textsubscript{BW}\approx2.42$ nm, which is in excellent agreement with the theory (see Supplement \S\ref{sec:Phasematching}).     %
We also observe that the Stokes sideband is 25 dB lower which confirms that the RF input produces an inter-polarization scattering process. Calibrations for these measurements are discussed in the Supplement \S\ref{sec:OpticalCharacterization}.

To further demonstrate the effect of the synthetic momentum bias, we also tested our devices with $\Delta\phi = -2\pi/3$ which reverses the direction of the RF stimulus. As an aside, this is equivalent to injecting TE carrier signal from the Port 2 while keeping the phase at $\Delta\phi = 2\pi/3$.
The modulation was measured near the strongest point (around 780 nm) and compared to the $\Delta\phi = 2\pi/3$ case with results summarized in Fig.~\ref{fig:4}c. We observe that the $\Delta\phi = -2\pi/3$ case flips the single sideband generation strongly towards the Stokes sideband as expected. This demonstrates an asymmetry in single sideband generation due to the directionality of the travelling wave, and confirms that the applied RF field breaks the time reversal symmetry of the optical waveguide leading to direction dependent mode conversion and frequency shift for the input light.
We also tested the power handling capability of the device, which is plotted for the anti-Stokes sideband ($\Delta\phi = 2\pi/3$ case) in Fig.~\ref{fig:4}d.
We observe that the polarization conversion rate is proportional to $\sqrt{P\textsubscript{RF}}$, i.e. proportional to voltage, as expected from any electro-optic modulator.
At the maximum applied RF power we observe $\approx 1.04 \%$ inter-polarization conversion over the 1.08 mm interaction length, corresponding to a polarization rotation rate of 0.94 rad/cm (see also Supplement \S{\ref{sec:OpticalCharacterization}}).
Similar to the Verdet constant that quantifies the magneto-optic polarization rotation normalized to the applied axial magnetic field, here we can divide by the applied voltage $\approx$ 7.1 V to obtain a normalized figure of merit $\approx$ 0.13 rad/$\rm{V}\cdot\rm{cm}$.

From this result, we additionally estimate the half-wave voltage product to be around 11.85 V$\cdot$cm, and the interaction length can be increased further to achieve full conversion (see Supplement~\S\ref{sec:OpticalCharacterization}).
As a point of comparison, the devices that have previously demonstrated \textit{reciprocal} polarization rotation in Ti-diffused thin-film LNOI waveguides have shown a best case half-wave voltage product of $\approx$ 30 V$\cdot$cm~\cite{Noe_Smith_1988,Campbell_Steier_1971} with a similar X-cut crystal. 
This highlights an advantage of the thin-film LNOI approach due to enhanced optical confinement in the nanophotonic waveguide which permits closer placement of the electrodes.
It is also instructive at this point to compare to the best MOFE results on chip, where figures of merit around 0.6 - 0.75 rad/dB have been estimated~\cite{Zhang:19,Yan:20} . Since in lithium niobate propagation losses in the range of 0.02 - 0.1 dB/cm have been reported \cite{Desiatov:19,Sohn_Orsel_Bahl_2021} we can estimate from the above 0.94 rad/cm result that the figure of merit for EO-NPRs can approach 10-50 rad/dB, which is significantly higher than MOFE devices.

\vspace{12pt} %

\begin{figure}[tp]
    \begin{adjustwidth*}{-1in}{-1in}
    \hsize=\linewidth
    \includegraphics[width=1.4\columnwidth]{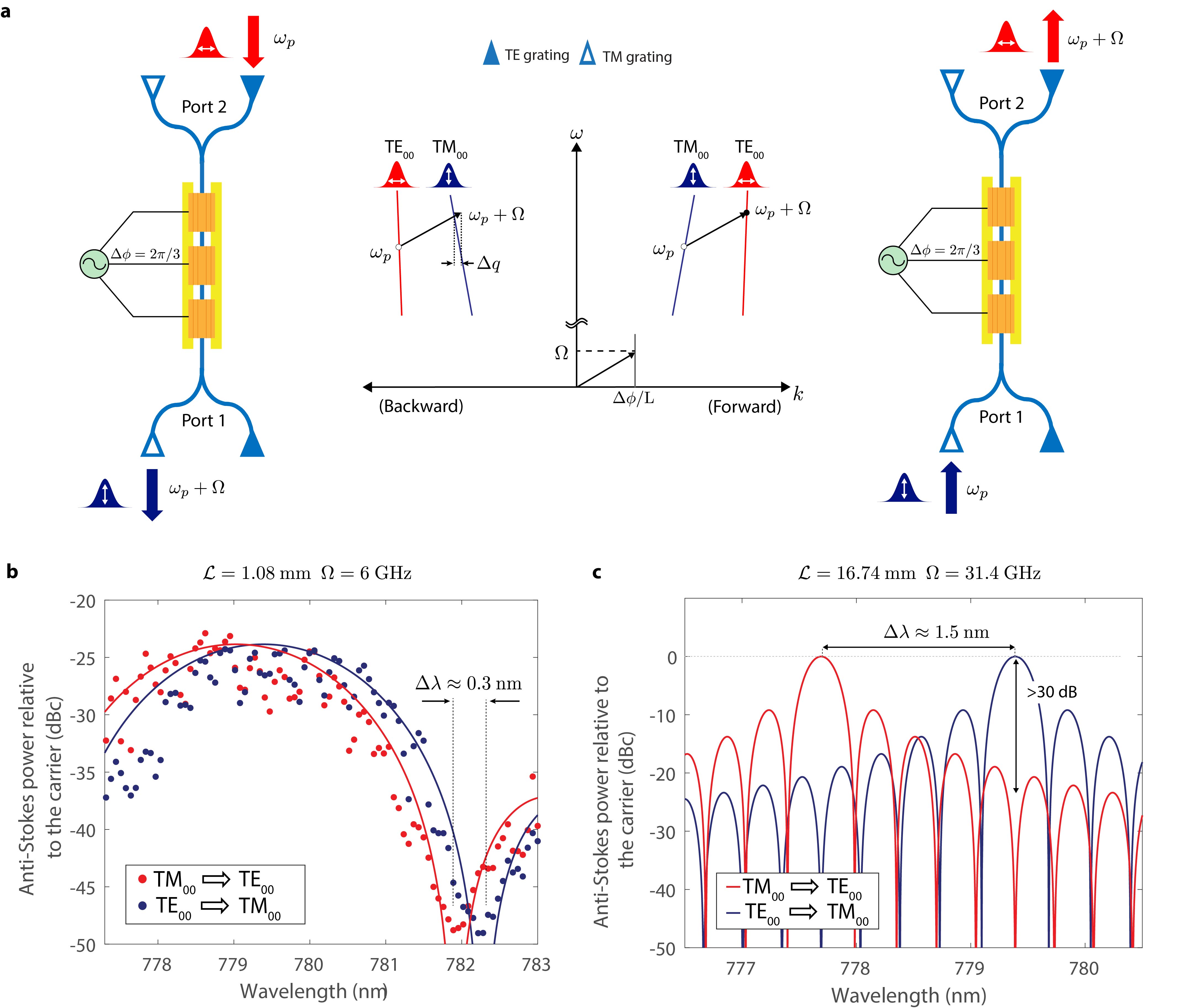}
    \centering
    \caption{
    \textbf{Experimental demonstration of non-reciprocal polarization rotation.}
    \textbf{(a)} 
        We consider anti-Stokes scattering for a probe of frequency $\omega_p$ with a fixed $\Delta\phi = 2\pi/3$. In the forward direction, with input on {\TMzz} the phase-matching condition is satisfied for scattering into the {\TEzz} mode. In the backward direction, anti-Stokes scattering is only satisfied if the input is on {\TEzz} and scattered into \TMzz. Even so, a non-reciprocal phase mismatch $\Delta q$ prohibits this process from ocurring at the same frequency in both directions.
    \textbf{(b)} Experimental results for a device with 1.08 mm length and 6 GHz RF modulation applied, with 21 dBm RF power applied to each electrode. The measurements confirm the spectrally-shifted sinc$^2$ responses and demonstrate non-reciprocal conversion. Solid lines are a fit to the theory.
    \textbf{(c)} Extrapolating from the experimental results, we can simulate a `full-conversion' EO-NPR with identical cross-section design, in which the interaction length is increased to $\mathcal{L}=16.74$ mm and the applied RF is increased to 31.4 GHz (discussion in Supplement \S\ref{sec:OpticalCharacterization}). The total RF power is unchanged at 21 dBm. The modulation frequency is selected to optimize the alignment between peaks and valleys in opposite directions so that contrast is maximized.
    }
    \label{fig:5}
    \end{adjustwidth*}
\end{figure}

The above experiment establishes asymmetry in \TMzz to \TEzz conversion depending on the direction of the travelling wave and demonstrates broken time-reversal symmetry. However this is not the typical mode of operation for a MOFE device.
Since the polarization is an important defining characteristic of what constitutes a `port', the more relevant case is that of an undesirable reflection coming back into the output mode (here \TEzz) as previously illustrated in Fig.~\ref{fig:1}e. 
For this we test the case illustrated in Fig.~\ref{fig:5}a, i.e. \TMzz to \TEzz in one direction and \TEzz to \TMzz in the opposite direction for a chosen stimulus configuration.
We note here that depending on the choice of $\Delta\phi$ for the RF stimulus, this results in either only Stokes or only anti-Stokes conversion for both inputs. For our demonstration, we tested the anti-Stokes conversion case to show the non-reciprocal polarization rotation.
As described earlier, the forward and backward sideband conversion processes should display a shifted sinc$^2$ response depending on the group index of the modes and the modulation frequency. 
To increase the relative shift between these sinc$^2$ functions and clearly show a strong non-reciprocal effect, we increased the RF frequency $\Omega$ to 6 GHz and set 21 dBm of applied RF power. This flexibility of the stimulus frequency is possible since the momentum shift is set by the electrode pitch $L$ and the relative phase shift $\Delta\phi$.
The resulting experimental observations are presented in Fig.~\ref{fig:5}b.
We observe a spectral shift of 0.3 nm between the sinc$^2$ conversion responses, with a peak sideband non-reciprocity of $\approx$ 9 dB, proving the non-reciprocal polarization rotation effect. This spectral shift is also in good agreement with our theoretical calculations (in Supplement \S\ref{sec:Nonreciprocal}) 
and can be increased by using larger RF frequencies as described earlier. Unfortunately, it was not possible to increase the frequency beyond 6 GHz in our current devices due to bandwidth limitations introduced by the electrodes (discussion in \S\ref{sec:Electricalmodel}).     
Presently, however, we can use the extracted data from Fig.~\ref{fig:5}b to extrapolate the operational characteristics of a device that could reach complete inter-polarization conversion with a longer interaction length $\mathcal{L} = 16.74$ mm (See Supplement \S\ref{sec:OpticalCharacterization}). This simulation is presented in Fig.~\ref{fig:5}c where we additionally set a higher RF frequency ($\Omega = $ 31.4 GHz) so that the conversion peaks and valleys are aligned for highest contrast.
\begin{figure}[tp]
    \begin{adjustwidth*}{-1in}{-1in}
    \hsize=\linewidth
    \includegraphics[width=1\columnwidth]{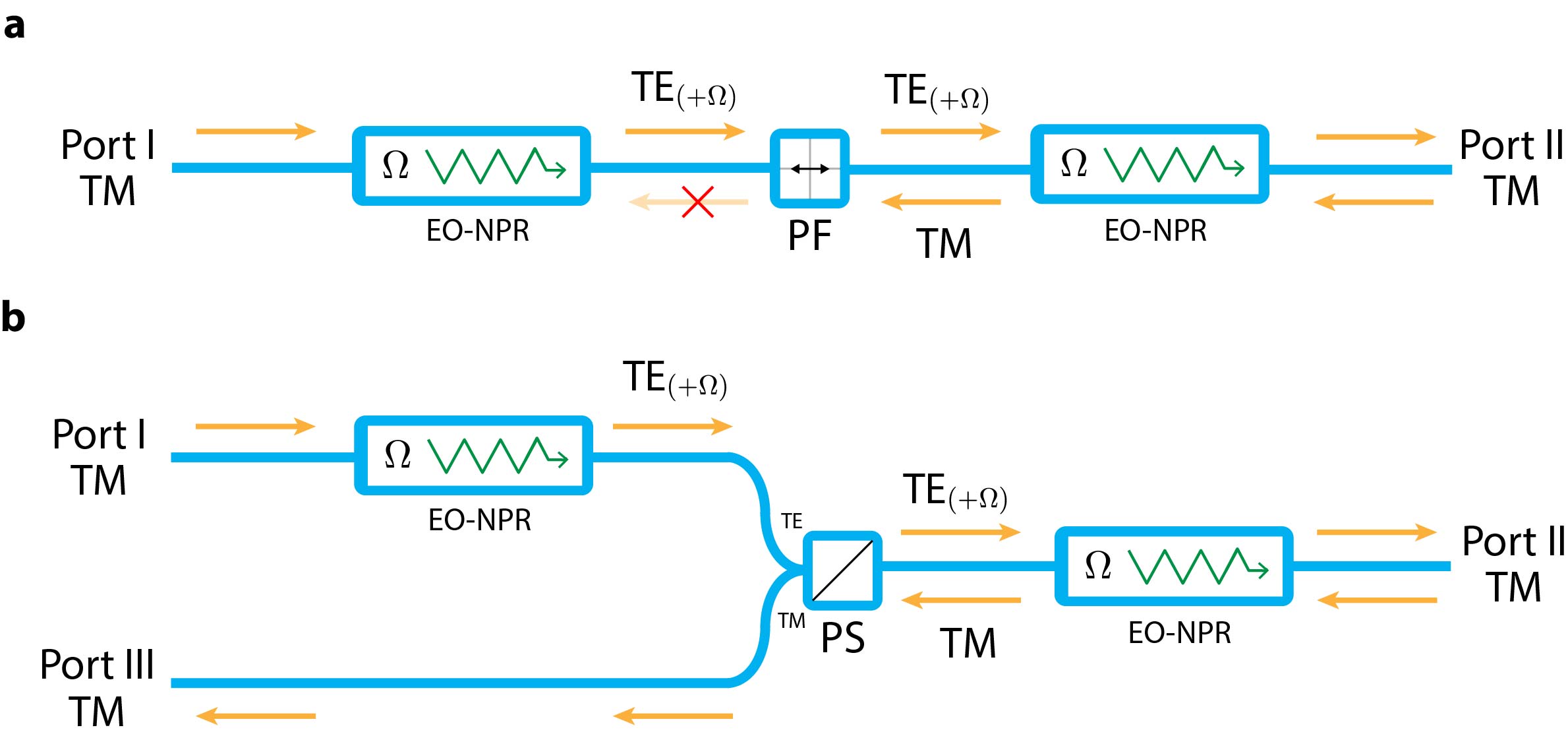}
    \centering
    \caption{
    \textbf{Potential designs of a 2-port isolator and 3-port circulator constructed with ``full conversion'' EO-NPRs from Fig.~\ref{fig:5}c.}
    \textbf{(a)} We can construct a linear non-frequency-shifting isolator by using two EO-NPRs with an intermediate polarization filter (PF).
    \textbf{(b)} A linear non-frequency shifting circulator can be implemented by replacing the PF in the isolator design with a polarization splitter (PS) that sends returning signals to a different output port.
    }
    \label{fig:6}
    \end{adjustwidth*}
\end{figure}

As a key point of distinction with the traditional MOFE, the electro-optic non-reciprocal polarization rotation that we have demonstrated is always accompanied by a fixed frequency shift of $\Omega$ that is set by the RF input tone.     
Even so, this EO-NPR effect can be leveraged in a manner similar to the MOFE by introducing appropriate polarization selective filters and splitters.
For instance, we can produce a linear optical isolator without any frequency shift, by using two full-conversion EO-NPRs with an intermediate polarization selective filter (Fig.~\ref{fig:6}a). As illustrated, in the forward direction the first EO-NPR converts TM input to a TE signal that passes through the filter and is subsequently converted back to TM. In the reverse direction the first EO-NPR does not perform conversion which allows the filter to reject the TM mode.
Linear 3-port circulator operation, again without any frequency shift, can also be achieved by replacing the intermediate polarization filter in the isolator design by a polarization splitter that leads to a drop port (Fig.~\ref{fig:6}b).

\vspace{12pt}  %

The ability to perform non-reciprocal polarization rotation has been generally absent from photonic integrated circuits, and yet, devices that leverage this effect are widely used in critical locations within photonic systems.
In this work we have demonstrated that an ultralow-loss magnetless alternative of the MOFE can be realized by leveraging electro-optic coefficients ($\rm{r}_{ijk, \, i\neq j}$) that are found in many photonic materials. This list includes active III-V platforms based on GaAs~{\cite{McCallum_Huang_Smirl_Sun_Towe_1995,Suzuki_Tada_1984}}, InP~{\cite{Suzuki_Tada_1984}}, and InGaP~{\cite{Ueno:97}}, potentially making it feasible to co-integrate lasers with high-bandwidth optical isolators and circulators without requiring heterogeneous integration.
Unlike magneto-optic isolators, this electro-optic approach also decouples the full-conversion interaction length from contrast control (the latter can be manipulated by the applied RF frequency) which can allow dynamic adjustment even after fabrication.
In addition, the operational wavelength range for EO-NPRs is set lithographically which makes short wavelength non-reciprocal devices much more accessible, especially for atom-photon integration and low-loss quantum photonics.
    
\vspace{24pt}

\section*{Methods}

\textbf{Fabrication -- }
The optical components (waveguides and grating couplers) were patterned using a 150 keV electron beam lithography (EBL) system and etched with an optimized Ar inductively coupled plasma reactive ion etch (ICP RIE) process~\cite{Sohn_Orsel_Bahl_2021}. The final ridge waveguide has approximately $60^o$ sidewall angle due to the physical etching process. The lower electrodes were similarly defined by EBL, and for the metallization, Au was deposited using e-beam evaporation followed by a lift-off process. A cladding layer of 1.5 um SiO$_2$ was then deposited by plasma enhanced chemical vapour deposition (PECVD) and metal via holes were defined in the cladding using oxide etching. Finally, another Au layer was deposited using e-beam evaporation and followed by lift-off process to pattern the top electrodes.

\vspace{12pt}

\noindent
\textbf{Measurement equipment -- }
The heterodyne measurement setup used in our experiments is presented and described in Fig.~\ref{fig:3}. We used a sub-50 kHz linewidth (770-783 nm) tunable external cavity diode laser (New Focus model TLB-6712-P) as the light source. 
An acousto-optic frequency shifter (Brimrose TEF-200-780-2FP) is used to generate a reference for frequency-resolved measurements of the carrier and sidebands using a high-speed photodetector (Thorlabs model RXM25DF). The probe direction is controlled by an off-chip optical switch (Thorlabs model OSW22-780E). For generating RF signals and measuring the optical beat spectrum, we used a 4-port vector network analyzer (Agilent ENA series E5080A). To produce the data in Fig.~\ref{fig:4}, the three RF stimulus phases were generated from the network analyzer output using adjustable RF phase shifters (Pasterneck PE8247). For producing the data in Fig.~\ref{fig:5} and Fig.~\ref{Electricalparameters}b, the RF signals were provided directly from a multi-output synchronized microwave signal generator (Holzworth 9004).

\vspace{24pt}

{\footnotesize \putbib}

{\footnotesize \bibliography{thesisrefsol}}

\begin{thebibliography}{10}
\expandafter\ifx\csname url\endcsname\relax
  \def\url#1{\texttt{#1}}\fi
\expandafter\ifx\csname urlprefix\endcsname\relax\def\urlprefix{URL }\fi
\providecommand{\bibinfo}[2]{#2}
\providecommand{\eprint}[2][]{\url{#2}}

\bibitem{Bennett_Stern_1965}
\bibinfo{author}{Bennett, H.~S.} \& \bibinfo{author}{Stern, E.~A.}
\newblock \bibinfo{title}{Faraday effect in solids}.
\newblock \emph{\bibinfo{journal}{Physical Review}}
  \textbf{\bibinfo{volume}{137}}, \bibinfo{pages}{A448–A461}
  (\bibinfo{year}{1965}).

\bibitem{Hulme_Fowler_1932}
\bibinfo{author}{Hulme, H.~R.} \& \bibinfo{author}{Fowler, R.~H.}
\newblock \bibinfo{title}{The faraday effect in ferromagnetics}.
\newblock \emph{\bibinfo{journal}{Proceedings of the Royal Society of London.
  Series A, Containing Papers of a Mathematical and Physical Character}}
  \textbf{\bibinfo{volume}{135}}, \bibinfo{pages}{237–257}
  (\bibinfo{year}{1932}).

\bibitem{Serber_1932}
\bibinfo{author}{Serber, R.}
\newblock \bibinfo{title}{The theory of the faraday effect in molecules}.
\newblock \emph{\bibinfo{journal}{Physical Review}}
  \textbf{\bibinfo{volume}{41}}, \bibinfo{pages}{489–506}
  (\bibinfo{year}{1932}).

\bibitem{Aplet_64}
\bibinfo{author}{Aplet, L.~J.} \& \bibinfo{author}{Carson, J.~W.}
\newblock \bibinfo{title}{A faraday effect optical isolator}.
\newblock \emph{\bibinfo{journal}{Appl. Opt.}} \textbf{\bibinfo{volume}{3}},
  \bibinfo{pages}{544--545} (\bibinfo{year}{1964}).

\bibitem{S_Fischer_1987}
\bibinfo{author}{Fischer, S.}
\newblock \bibinfo{title}{The faraday optical isolator}.
\newblock \emph{\bibinfo{journal}{Journal of Optical Communications}}
  \textbf{\bibinfo{volume}{8}}, \bibinfo{pages}{18–21}
  (\bibinfo{year}{1987}).

\bibitem{Wolfe_Hegarty_Dillon_Luther_Celler_Trimble_Dorsey_1985}
\bibinfo{author}{Wolfe, R.} \emph{et~al.}
\newblock \bibinfo{title}{Thin‐film waveguide magneto‐optic isolator}.
\newblock \emph{\bibinfo{journal}{Applied Physics Letters}}
  \textbf{\bibinfo{volume}{46}}, \bibinfo{pages}{817–819}
  (\bibinfo{year}{1985}).

\bibitem{Zhang:19}
\bibinfo{author}{Zhang, Y.} \emph{et~al.}
\newblock \bibinfo{title}{Monolithic integration of broadband optical isolators
  for polarization-diverse silicon photonics}.
\newblock \emph{\bibinfo{journal}{Optica}} \textbf{\bibinfo{volume}{6}},
  \bibinfo{pages}{473--478} (\bibinfo{year}{2019}).

\bibitem{Yan:20}
\bibinfo{author}{Yan, W.} \emph{et~al.}
\newblock \bibinfo{title}{Waveguide-integrated high-performance magneto-optical
  isolators and circulators on silicon nitride platforms}.
\newblock \emph{\bibinfo{journal}{Optica}} \textbf{\bibinfo{volume}{7}},
  \bibinfo{pages}{1555--1562} (\bibinfo{year}{2020}).

\bibitem{Srinivasan_Stadler_2018}
\bibinfo{author}{Srinivasan, K.} \& \bibinfo{author}{Stadler, B. J.~H.}
\newblock \bibinfo{title}{Magneto-optical materials and designs for integrated
  {TE} - and {TM} - mode planar waveguide isolators: a review [invited]}.
\newblock \emph{\bibinfo{journal}{Optical Materials Express}}
  \textbf{\bibinfo{volume}{8}}, \bibinfo{pages}{3307–3318}
  (\bibinfo{year}{2018}).

\bibitem{Ross:11}
\bibinfo{author}{Bi, L.} \emph{et~al.}
\newblock \bibinfo{title}{On-chip optical isolation in monolithically
  integrated non-reciprocal optical resonators}.
\newblock \emph{\bibinfo{journal}{Nat. Photonics}}
  \textbf{\bibinfo{volume}{5}}, \bibinfo{pages}{758} (\bibinfo{year}{2011}).

\bibitem{Ghosh:12aa}
\bibinfo{author}{Ghosh, S.} \emph{et~al.}
\newblock \bibinfo{title}{{Ce:YIG/Silicon}-on-insulator waveguide optical
  isolator realized by adhesive bonding}.
\newblock \emph{\bibinfo{journal}{Opt. Express}} \textbf{\bibinfo{volume}{20}},
  \bibinfo{pages}{1839--1848} (\bibinfo{year}{2012}).

\bibitem{Huang:17}
\bibinfo{author}{Huang, D.} \emph{et~al.}
\newblock \bibinfo{title}{Dynamically reconfigurable integrated optical
  circulators}.
\newblock \emph{\bibinfo{journal}{Optica}} \textbf{\bibinfo{volume}{4}},
  \bibinfo{pages}{23--30} (\bibinfo{year}{2017}).

\bibitem{Zhang:2017wq}
\bibinfo{author}{Zhang, C.}, \bibinfo{author}{Dulal, P.},
  \bibinfo{author}{Stadler, B. J.~H.} \& \bibinfo{author}{Hutchings, D.~C.}
\newblock \bibinfo{title}{Monolithically-integrated {TE}-mode {1D}
  silicon-on-insulator isolators using seedlayer-free garnet}.
\newblock \emph{\bibinfo{journal}{Scientific Reports}}
  \textbf{\bibinfo{volume}{7}}, \bibinfo{pages}{5820} (\bibinfo{year}{2017}).

\bibitem{Du:2018wy}
\bibinfo{author}{Du, Q.} \emph{et~al.}
\newblock \bibinfo{title}{Monolithic on-chip magneto-optical isolator with 3
  {dB} insertion loss and 40 db isolation ratio}.
\newblock \emph{\bibinfo{journal}{ACS Photonics}} \textbf{\bibinfo{volume}{5}},
  \bibinfo{pages}{5010--5016} (\bibinfo{year}{2018}).

\bibitem{Ruesink:16}
\bibinfo{author}{Ruesink, F.}, \bibinfo{author}{Miri, M.-A.},
  \bibinfo{author}{Al{\`u}, A.} \& \bibinfo{author}{Verhagen, E.}
\newblock \bibinfo{title}{Nonreciprocity and magnetic-free isolation based on
  optomechanical interactions}.
\newblock \emph{\bibinfo{journal}{Nat. Commun.}} \textbf{\bibinfo{volume}{7}},
  \bibinfo{pages}{13662} (\bibinfo{year}{2016}).

\bibitem{Sohn18}
\bibinfo{author}{Sohn, D.~B.}, \bibinfo{author}{Kim, S.} \&
  \bibinfo{author}{Bahl, G.}
\newblock \bibinfo{title}{Time-reversal symmetry breaking with acoustic pumping
  of nanophotonic circuits}.
\newblock \emph{\bibinfo{journal}{Nat. Photonics}}
  \textbf{\bibinfo{volume}{12}}, \bibinfo{pages}{91--97}
  (\bibinfo{year}{2018}).

\bibitem{Kittlaus_2018}
\bibinfo{author}{Kittlaus, E.~A.}, \bibinfo{author}{Otterstrom, N.~T.},
  \bibinfo{author}{Kharel, P.}, \bibinfo{author}{Gertler, S.} \&
  \bibinfo{author}{Rakich, P.~T.}
\newblock \bibinfo{title}{Non-reciprocal interband brillouin modulation}.
\newblock \emph{\bibinfo{journal}{Nature Photonics}}
  \textbf{\bibinfo{volume}{12}}, \bibinfo{pages}{613–619}
  (\bibinfo{year}{2018}).

\bibitem{Sohn:2019aa}
\bibinfo{author}{Sohn, D.~B.} \& \bibinfo{author}{Bahl, G.}
\newblock \bibinfo{title}{Direction reconfigurable nonreciprocal acousto-optic
  modulator on chip}.
\newblock \emph{\bibinfo{journal}{APL Photonics}} \textbf{\bibinfo{volume}{4}},
  \bibinfo{pages}{126103} (\bibinfo{year}{2019}).

\bibitem{Tian:2020ti}
\bibinfo{author}{Tian, H.} \emph{et~al.}
\newblock \bibinfo{title}{Magnetic-free silicon nitride integrated optical
  isolator}.
\newblock \emph{\bibinfo{journal}{Nature Photonics}}
  \textbf{\bibinfo{volume}{15}}, \bibinfo{pages}{828–836}
  (\bibinfo{year}{2021}).

\bibitem{sarabalis2020}
\bibinfo{author}{Sarabalis, C.~J.} \emph{et~al.}
\newblock \bibinfo{title}{Acousto-optic modulation of a wavelength-scale
  waveguide}.
\newblock \emph{\bibinfo{journal}{Optica}} \textbf{\bibinfo{volume}{8}},
  \bibinfo{pages}{477--483} (\bibinfo{year}{2021}).

\bibitem{Kittlaus:2021wq}
\bibinfo{author}{Kittlaus, E.~A.} \emph{et~al.}
\newblock \bibinfo{title}{Electrically driven acousto-optics and broadband
  non-reciprocity in silicon photonics}.
\newblock \emph{\bibinfo{journal}{Nature Photonics}}
  \textbf{\bibinfo{volume}{15}}, \bibinfo{pages}{43--52}
  (\bibinfo{year}{2021}).

\bibitem{Sohn_Orsel_Bahl_2021}
\bibinfo{author}{Sohn, D.~B.}, \bibinfo{author}{Örsel, O.~E.} \&
  \bibinfo{author}{Bahl, G.}
\newblock \bibinfo{title}{Electrically driven optical isolation through
  phonon-mediated photonic autler–townes splitting}.
\newblock \emph{\bibinfo{journal}{Nature Photonics}}
  \textbf{\bibinfo{volume}{15}}, \bibinfo{pages}{822–827}
  (\bibinfo{year}{2021}).

\bibitem{Sounas:14}
\bibinfo{author}{Sounas, D.~L.} \& \bibinfo{author}{Al{\`u}, A.}
\newblock \bibinfo{title}{Angular-momentum-biased nanorings to realize
  magnetic-free integrated optical isolation}.
\newblock \emph{\bibinfo{journal}{ACS Photonics}} \textbf{\bibinfo{volume}{1}},
  \bibinfo{pages}{198--204} (\bibinfo{year}{2014}).

\bibitem{Doerr:11}
\bibinfo{author}{Doerr, C.~R.}, \bibinfo{author}{Dupuis, N.} \&
  \bibinfo{author}{Zhang, L.}
\newblock \bibinfo{title}{Optical isolator using two tandem phase modulators}.
\newblock \emph{\bibinfo{journal}{Opt. Lett.}} \textbf{\bibinfo{volume}{36}},
  \bibinfo{pages}{4293--4295} (\bibinfo{year}{2011}).

\bibitem{lira2012}
\bibinfo{author}{Lira, H.}, \bibinfo{author}{Yu, Z.}, \bibinfo{author}{Fan, S.}
  \& \bibinfo{author}{Lipson, M.}
\newblock \bibinfo{title}{Electrically driven nonreciprocity induced by
  interband photonic transition on a silicon chip}.
\newblock \emph{\bibinfo{journal}{Phys. Rev. Lett.}}
  \textbf{\bibinfo{volume}{109}}, \bibinfo{pages}{033901}
  (\bibinfo{year}{2012}).

\bibitem{Tzuang2014}
\bibinfo{author}{Tzuang, L.~D.}, \bibinfo{author}{Fang, K.},
  \bibinfo{author}{Nussenzveig, P.}, \bibinfo{author}{Fan, S.} \&
  \bibinfo{author}{Lipson, M.}
\newblock \bibinfo{title}{Non-reciprocal phase shift induced by an effective
  magnetic flux for light}.
\newblock \emph{\bibinfo{journal}{Nat. Photonics}}
  \textbf{\bibinfo{volume}{8}}, \bibinfo{pages}{701} (\bibinfo{year}{2014}).

\bibitem{Li:2014vo}
\bibinfo{author}{Li, E.}, \bibinfo{author}{Eggleton, B.~J.},
  \bibinfo{author}{Fang, K.} \& \bibinfo{author}{Fan, S.}
\newblock \bibinfo{title}{Photonic aharonov--bohm effect in photon--phonon
  interactions}.
\newblock \emph{\bibinfo{journal}{Nature Communications}}
  \textbf{\bibinfo{volume}{5}}, \bibinfo{pages}{3225} (\bibinfo{year}{2014}).

\bibitem{Dostart:2021uh}
\bibinfo{author}{Dostart, N.}, \bibinfo{author}{Gevorgyan, H.},
  \bibinfo{author}{Onural, D.} \& \bibinfo{author}{Popovi{\'c}, M.}
\newblock \bibinfo{title}{Optical isolation using microring modulators}.
\newblock \emph{\bibinfo{journal}{Optics Letters}}
  \textbf{\bibinfo{volume}{46}}, \bibinfo{pages}{460--463}
  (\bibinfo{year}{2021}).

\bibitem{Desiatov:19}
\bibinfo{author}{Desiatov, B.}, \bibinfo{author}{Shams-Ansari, A.},
  \bibinfo{author}{Zhang, M.}, \bibinfo{author}{Wang, C.} \&
  \bibinfo{author}{Lon\v{c}ar, M.}
\newblock \bibinfo{title}{Ultra-low-loss integrated visible photonics using
  thin-film lithium niobate}.
\newblock \emph{\bibinfo{journal}{Optica}} \textbf{\bibinfo{volume}{6}},
  \bibinfo{pages}{380--384} (\bibinfo{year}{2019}).

\bibitem{Wang_2018}
\bibinfo{author}{Wang, C.} \emph{et~al.}
\newblock \bibinfo{title}{Integrated lithium niobate electro-optic modulators
  operating at cmos-compatible voltages}.
\newblock \emph{\bibinfo{journal}{Nature}} \textbf{\bibinfo{volume}{562}},
  \bibinfo{pages}{101–104} (\bibinfo{year}{2018}).

\bibitem{Zhang_comb_2019}
\bibinfo{author}{Zhang, M.} \emph{et~al.}
\newblock \bibinfo{title}{Broadband electro-optic frequency comb generation in
  a lithium niobate microring resonator}.
\newblock \emph{\bibinfo{journal}{Nature}} \textbf{\bibinfo{volume}{568}},
  \bibinfo{pages}{373–377} (\bibinfo{year}{2019}).

\bibitem{Hu_Nat_2021}
\bibinfo{author}{Hu, Y.} \emph{et~al.}
\newblock \bibinfo{title}{On-chip electro-optic frequency shifters and beam
  splitters}.
\newblock \emph{\bibinfo{journal}{Nature}} \textbf{\bibinfo{volume}{599}},
  \bibinfo{pages}{587–593} (\bibinfo{year}{2021}).

\bibitem{Zhu:21}
\bibinfo{author}{Zhu, D.} \emph{et~al.}
\newblock \bibinfo{title}{Integrated photonics on thin-film lithium niobate}.
\newblock \emph{\bibinfo{journal}{Adv. Opt. Photon.}}
  \textbf{\bibinfo{volume}{13}}, \bibinfo{pages}{242--352}
  (\bibinfo{year}{2021}).

\bibitem{Campbell_Steier_1971}
\bibinfo{author}{Campbell, J.} \& \bibinfo{author}{Steier, W.}
\newblock \bibinfo{title}{Rotating-waveplate optical-frequency shifting in
  lithium niobate}.
\newblock \emph{\bibinfo{journal}{IEEE Journal of Quantum Electronics}}
  \textbf{\bibinfo{volume}{7}}, \bibinfo{pages}{450–457}
  (\bibinfo{year}{1971}).

\bibitem{Noe_Smith_1988}
\bibinfo{author}{Noe, R.} \& \bibinfo{author}{Smith, D.~A.}
\newblock \bibinfo{title}{Integrated-optic rotating waveplate frequency
  shifter}.
\newblock \emph{\bibinfo{journal}{Electronics Letters}}
  \textbf{\bibinfo{volume}{24}}, \bibinfo{pages}{1348–1349}
  (\bibinfo{year}{1988}).

\bibitem{Qin_Lu_Pollick_Sriram_Yoo_2017}
\bibinfo{author}{Qin, C.}, \bibinfo{author}{Lu, H.}, \bibinfo{author}{Pollick,
  A.}, \bibinfo{author}{Sriram, S.} \& \bibinfo{author}{Yoo, S. J.~B.}
\newblock \bibinfo{title}{Power-efficient electro-optical single-tone
  optical-frequency shifter using x-cut y-propagating lithium tantalate
  waveguide emulating a rotating half-wave-plate}.
\newblock In \emph{\bibinfo{booktitle}{Optical Fiber Communication Conference
  (2017), paper Th3I.4}}, \bibinfo{pages}{Th3I.4} (\bibinfo{publisher}{Optica
  Publishing Group}, \bibinfo{year}{2017}).

\bibitem{Majumder_Shen_Polson_Menon_2017}
\bibinfo{author}{Majumder, A.}, \bibinfo{author}{Shen, B.},
  \bibinfo{author}{Polson, R.} \& \bibinfo{author}{Menon, R.}
\newblock \bibinfo{title}{Ultra-compact polarization rotation in integrated
  silicon photonics using digital metamaterials}.
\newblock \emph{\bibinfo{journal}{Optics Express}}
  \textbf{\bibinfo{volume}{25}}, \bibinfo{pages}{19721–19731}
  (\bibinfo{year}{2017}).

\bibitem{Posner_Podoliak_Smith_Mennea_Horak_Gawith_Smith_Gates_2019}
\bibinfo{author}{Posner, M.~T.} \emph{et~al.}
\newblock \bibinfo{title}{Integrated polarizer based on 45° tilted gratings}.
\newblock \emph{\bibinfo{journal}{Optics Express}}
  \textbf{\bibinfo{volume}{27}}, \bibinfo{pages}{11174–11181}
  (\bibinfo{year}{2019}).

\bibitem{Hwang:97}
\bibinfo{author}{Hwang, I.~K.}, \bibinfo{author}{Yun, S.~H.} \&
  \bibinfo{author}{Kim, B.~Y.}
\newblock \bibinfo{title}{All-fiber-optic nonreciprocal modulator}.
\newblock \emph{\bibinfo{journal}{Opt. Lett.}} \textbf{\bibinfo{volume}{22}},
  \bibinfo{pages}{507--509} (\bibinfo{year}{1997}).

\bibitem{Yu_Fan_2009}
\bibinfo{author}{Yu, Z.} \& \bibinfo{author}{Fan, S.}
\newblock \bibinfo{title}{Complete optical isolation created by indirect
  interband photonic transitions}.
\newblock \emph{\bibinfo{journal}{Nature Photonics}}
  \textbf{\bibinfo{volume}{3}}, \bibinfo{pages}{91–94}
  (\bibinfo{year}{2009}).

\bibitem{Kim2015}
\bibinfo{author}{Kim, J.}, \bibinfo{author}{Kuzyk, M.~C.},
  \bibinfo{author}{Han, K.}, \bibinfo{author}{Wang, H.} \&
  \bibinfo{author}{Bahl, G.}
\newblock \bibinfo{title}{{Non-reciprocal Brillouin scattering induced
  transparency}}.
\newblock \emph{\bibinfo{journal}{Nat. Phys}} \textbf{\bibinfo{volume}{11}},
  \bibinfo{pages}{275--280} (\bibinfo{year}{2015}).

\bibitem{Huang_Lin_Chen_Huang_2007}
\bibinfo{author}{Huang, C.~Y.}, \bibinfo{author}{Lin, C.~H.},
  \bibinfo{author}{Chen, Y.~H.} \& \bibinfo{author}{Huang, Y.~C.}
\newblock \bibinfo{title}{Electro-optic ti:ppln waveguide as efficient optical
  wavelength filter and polarization mode converter}.
\newblock \emph{\bibinfo{journal}{Optics Express}}
  \textbf{\bibinfo{volume}{15}}, \bibinfo{pages}{2548–2554}
  (\bibinfo{year}{2007}).

\bibitem{Ding_Zheng_Chen_2019}
\bibinfo{author}{Ding, T.}, \bibinfo{author}{Zheng, Y.} \&
  \bibinfo{author}{Chen, X.}
\newblock \bibinfo{title}{On-chip solc-type polarization control and wavelength
  filtering utilizing periodically poled lithium niobate on insulator ridge
  waveguide}.
\newblock \emph{\bibinfo{journal}{Journal of Lightwave Technology}}
  \textbf{\bibinfo{volume}{37}}, \bibinfo{pages}{1296–1300}
  (\bibinfo{year}{2019}).

\bibitem{Weis:1985aa}
\bibinfo{author}{Weis, R.~S.} \& \bibinfo{author}{Gaylord, T.~K.}
\newblock \bibinfo{title}{Lithium niobate: Summary of physical properties and
  crystal structure}.
\newblock \emph{\bibinfo{journal}{Applied Physics A}}
  \textbf{\bibinfo{volume}{37}}, \bibinfo{pages}{191--203}
  (\bibinfo{year}{1985}).

\bibitem{Boyd}
\bibinfo{author}{Boyd, R.~W.}
\newblock \emph{\bibinfo{title}{Nonlinear Optics, Third Edition}}
  (\bibinfo{publisher}{Academic Press, Inc.}, \bibinfo{address}{USA},
  \bibinfo{year}{2008}), \bibinfo{edition}{3rd} edn.

\bibitem{McCallum_Huang_Smirl_Sun_Towe_1995}
\bibinfo{author}{McCallum, D.~S.}, \bibinfo{author}{Huang, X.~R.},
  \bibinfo{author}{Smirl, A.~L.}, \bibinfo{author}{Sun, D.} \&
  \bibinfo{author}{Towe, E.}
\newblock \bibinfo{title}{Polarization rotation modulator in a strained
  [110]‐oriented multiple quantum well}.
\newblock \emph{\bibinfo{journal}{Applied Physics Letters}}
  \textbf{\bibinfo{volume}{66}}, \bibinfo{pages}{2885–2887}
  (\bibinfo{year}{1995}).

\bibitem{Suzuki_Tada_1984}
\bibinfo{author}{Suzuki, N.} \& \bibinfo{author}{Tada, K.}
\newblock \bibinfo{title}{Electrooptic properties and raman scattering in
  {InP}}.
\newblock \emph{\bibinfo{journal}{Japanese Journal of Applied Physics}}
  \textbf{\bibinfo{volume}{23}}, \bibinfo{pages}{291} (\bibinfo{year}{1984}).

\bibitem{Ueno:97}
\bibinfo{author}{Ueno, Y.}, \bibinfo{author}{Ricci, V.} \&
  \bibinfo{author}{Stegeman, G.~I.}
\newblock \bibinfo{title}{Second-order susceptibility of
  {$\textrm{Ga}_{0.5}\textrm{In}_{0.5}\textrm{P}$} crystals at 1.5 {\textmu}m
  and their feasibility for waveguide quasi-phase matching}.
\newblock \emph{\bibinfo{journal}{J. Opt. Soc. Am. B}}
  \textbf{\bibinfo{volume}{14}}, \bibinfo{pages}{1428--1436}
  (\bibinfo{year}{1997}).

\bibitem{Ghione_2009}
\bibinfo{author}{Ghione, G.}
\newblock \emph{\bibinfo{title}{Semiconductor devices for high-speed
  optoelectronics}} (\bibinfo{publisher}{Cambridge University Press},
  \bibinfo{year}{2009}).

\bibitem{Kim:2021te}
\bibinfo{author}{Kim, S.}, \bibinfo{author}{Sohn, D.~B.},
  \bibinfo{author}{Peterson, C.~W.} \& \bibinfo{author}{Bahl, G.}
\newblock \bibinfo{title}{On-chip optical non-reciprocity through a synthetic
  hall effect for photons}.
\newblock \emph{\bibinfo{journal}{APL Photonics}} \textbf{\bibinfo{volume}{6}},
  \bibinfo{pages}{011301} (\bibinfo{year}{2021}).

\end{thebibliography}


\begin{thebibliography}{10}
\expandafter\ifx\csname url\endcsname\relax
  \def\url#1{\texttt{#1}}\fi
\expandafter\ifx\csname urlprefix\endcsname\relax\def\urlprefix{URL }\fi
\providecommand{\bibinfo}[2]{#2}
\providecommand{\eprint}[2][]{\url{#2}}

\bibitem{Weis:1985aa}
\bibinfo{author}{Weis, R.~S.} \& \bibinfo{author}{Gaylord, T.~K.}
\newblock \bibinfo{title}{Lithium niobate: Summary of physical properties and
  crystal structure}.
\newblock \emph{\bibinfo{journal}{Applied Physics A}}
  \textbf{\bibinfo{volume}{37}}, \bibinfo{pages}{191--203}
  (\bibinfo{year}{1985}).

\bibitem{Balram:14}
\bibinfo{author}{Balram, K.~C.}, \bibinfo{author}{Davan\c{c}o, M.},
  \bibinfo{author}{Lim, J.~Y.}, \bibinfo{author}{Song, J.~D.} \&
  \bibinfo{author}{Srinivasan, K.}
\newblock \bibinfo{title}{Moving boundary and photoelastic coupling in gaas
  optomechanical resonators}.
\newblock \emph{\bibinfo{journal}{Optica}} \textbf{\bibinfo{volume}{1}},
  \bibinfo{pages}{414--420} (\bibinfo{year}{2014}).

\bibitem{Kharel_2016}
\bibinfo{author}{Kharel, P.}, \bibinfo{author}{Behunin, R.~O.},
  \bibinfo{author}{Renninger, W.~H.} \& \bibinfo{author}{Rakich, P.~T.}
\newblock \bibinfo{title}{Noise and dynamics in forward brillouin
  interactions}.
\newblock \emph{\bibinfo{journal}{Physical Review A}}
  \textbf{\bibinfo{volume}{93}}, \bibinfo{pages}{063806}
  (\bibinfo{year}{2016}).

\bibitem{Kittlaus_2018}
\bibinfo{author}{Kittlaus, E.~A.}, \bibinfo{author}{Otterstrom, N.~T.},
  \bibinfo{author}{Kharel, P.}, \bibinfo{author}{Gertler, S.} \&
  \bibinfo{author}{Rakich, P.~T.}
\newblock \bibinfo{title}{Non-reciprocal interband brillouin modulation}.
\newblock \emph{\bibinfo{journal}{Nature Photonics}}
  \textbf{\bibinfo{volume}{12}}, \bibinfo{pages}{613–619}
  (\bibinfo{year}{2018}).

\bibitem{Yariv_1973}
\bibinfo{author}{Yariv, A.}
\newblock \bibinfo{title}{Coupled-mode theory for guided-wave optics}.
\newblock \emph{\bibinfo{journal}{IEEE Journal of Quantum Electronics}}
  \textbf{\bibinfo{volume}{9}}, \bibinfo{pages}{919–933}
  (\bibinfo{year}{1973}).

\bibitem{Sounas:14}
\bibinfo{author}{Sounas, D.~L.} \& \bibinfo{author}{Al{\`u}, A.}
\newblock \bibinfo{title}{Angular-momentum-biased nanorings to realize
  magnetic-free integrated optical isolation}.
\newblock \emph{\bibinfo{journal}{ACS Photonics}} \textbf{\bibinfo{volume}{1}},
  \bibinfo{pages}{198--204} (\bibinfo{year}{2014}).

\bibitem{Oppenheim_signal}
\bibinfo{author}{Oppenheim, A.~V.}, \bibinfo{author}{Willsky, A.~S.} \&
  \bibinfo{author}{Nawab, S.~H.}
\newblock \emph{\bibinfo{title}{Signals $\&$ Systems (2nd Ed.)}}
  (\bibinfo{publisher}{Prentice-Hall, Inc.}, \bibinfo{address}{USA},
  \bibinfo{year}{1996}).

\bibitem{Ghione_2009}
\bibinfo{author}{Ghione, G.}
\newblock \emph{\bibinfo{title}{Semiconductor devices for high-speed
  optoelectronics}} (\bibinfo{publisher}{Cambridge University Press},
  \bibinfo{year}{2009}).

\bibitem{Wang_2018}
\bibinfo{author}{Wang, C.} \emph{et~al.}
\newblock \bibinfo{title}{Integrated lithium niobate electro-optic modulators
  operating at cmos-compatible voltages}.
\newblock \emph{\bibinfo{journal}{Nature}} \textbf{\bibinfo{volume}{562}},
  \bibinfo{pages}{101–104} (\bibinfo{year}{2018}).

\bibitem{Zhu:21}
\bibinfo{author}{Zhu, D.} \emph{et~al.}
\newblock \bibinfo{title}{Integrated photonics on thin-film lithium niobate}.
\newblock \emph{\bibinfo{journal}{Adv. Opt. Photon.}}
  \textbf{\bibinfo{volume}{13}}, \bibinfo{pages}{242--352}
  (\bibinfo{year}{2021}).

\bibitem{Kim_Lee_Yun_2012}
\bibinfo{author}{Kim, M.-G.}, \bibinfo{author}{Lee, B.~H.} \&
  \bibinfo{author}{Yun, T.-Y.}
\newblock \bibinfo{title}{Equivalent-circuit model for high-capacitance mlcc
  based on transmission-line theory}.
\newblock \emph{\bibinfo{journal}{IEEE Transactions on Components, Packaging
  and Manufacturing Technology}} \textbf{\bibinfo{volume}{2}},
  \bibinfo{pages}{1012–1020} (\bibinfo{year}{2012}).

\bibitem{Zhang:19}
\bibinfo{author}{Zhang, Y.} \emph{et~al.}
\newblock \bibinfo{title}{Monolithic integration of broadband optical isolators
  for polarization-diverse silicon photonics}.
\newblock \emph{\bibinfo{journal}{Optica}} \textbf{\bibinfo{volume}{6}},
  \bibinfo{pages}{473--478} (\bibinfo{year}{2019}).

\bibitem{Yan:20}
\bibinfo{author}{Yan, W.} \emph{et~al.}
\newblock \bibinfo{title}{Waveguide-integrated high-performance magneto-optical
  isolators and circulators on silicon nitride platforms}.
\newblock \emph{\bibinfo{journal}{Optica}} \textbf{\bibinfo{volume}{7}},
  \bibinfo{pages}{1555--1562} (\bibinfo{year}{2020}).

\bibitem{Sohn_Orsel_Bahl_2021}
\bibinfo{author}{Sohn, D.~B.}, \bibinfo{author}{Örsel, O.~E.} \&
  \bibinfo{author}{Bahl, G.}
\newblock \bibinfo{title}{Electrically driven optical isolation through
  phonon-mediated photonic autler–townes splitting}.
\newblock \emph{\bibinfo{journal}{Nature Photonics}}
  \textbf{\bibinfo{volume}{15}}, \bibinfo{pages}{822–827}
  (\bibinfo{year}{2021}).

\end{thebibliography}


\begin{thebibliography}{10}
\expandafter\ifx\csname url\endcsname\relax
  \def\url#1{\texttt{#1}}\fi
\expandafter\ifx\csname urlprefix\endcsname\relax\def\urlprefix{URL }\fi
\providecommand{\bibinfo}[2]{#2}
\providecommand{\eprint}[2][]{\url{#2}}

\end{thebibliography}
\end{bibunit}
\section*{Acknowledgments}

This work was sponsored by the Defense Advanced Research Projects Agency (DARPA) grant FA8650-19-2-7924 and the Air Force Office of Scientific Research (AFOSR) grant FA9550-19-1-0256. GB would additionally like to acknowledge support from the Office of Naval Research (ONR) Director for Research Early Career grant N00014-17-1-2209 and the Presidential Early Career Award for Scientists and Engineers.

\FloatBarrier

\newpage

\renewcommand*{\citenumfont}[1]{S#1}
\renewcommand*{\bibnumfmt}[1]{[S#1]}
\newcommand{\beginsupplement}{%
        \setcounter{table}{0}
        \renewcommand{\thetable}{S\arabic{table}}%
        \setcounter{figure}{0}
        \renewcommand{\thefigure}{S\arabic{figure}}%
        \setcounter{equation}{0}
        \renewcommand{\theequation}{S\arabic{equation}}%
        \setcounter{section}{0}
        \renewcommand{\thesection}{S\arabic{section}}%
}

\beginsupplement
\begin{bibunit}

\begin{center}

\Large{\textbf{Supplementary Information: \\ Electro-optic non-reciprocal polarization rotation in lithium niobate}} \\
\vspace{12pt}
\vspace{12pt}
\large{
{Oğulcan E. Örsel $^1$},
and {Gaurav Bahl $^2$}} \\
\vspace{12pt}
    \footnotesize{$^1$ Department of Electrical $\&$ Computer Engineering,} \\
    \footnotesize{$^2$ Department of Mechanical Science and Engineering,} \\
    \footnotesize{University of Illinois at Urbana–Champaign, Urbana, IL 61801 USA} \\
\end{center}

\vspace{24pt}

\section{Calculation of the polarization rotation rate}
\label{sec:EOrate}
\setcounter{page}{1}

In general, any three-wave mixing process that leverages a perturbation of the dielectric permittivity must satisfy the phase matching requirements:

\begin{equation}
\hbar\omega_2=\hbar\omega_1\pm\hbar\Omega
\end{equation}

\begin{equation}
\hbar \vec{k_2}=\hbar \vec{k_1}\pm\hbar \Vec{q}
\end{equation}

\noindent
where $\omega_1$ and $\omega_2$ represent the optical frequencies, $\Omega$ represents the frequency of the perturbing field, $\vec{k_1}$ and $\vec{k_2}$ represent the momentum of the photons, and $\vec{q}$ represents the momentum of the perturbing field. 
In order to simplify the analysis, suitable for our experimental case, we consider optical modes that are propagating along a 1D waveguide in the z-direction as: 

\begin{equation}
\textbf{E}^\textrm{TM}(\boldsymbol{\mathfrak{r}})=a_1(z) \, \boldsymbol{\mathcal{E}^\textrm{TM}(\boldsymbol{\mathfrak{r}_\bot})} \,e^{i(k_1 z-\omega_1 t)}+c.c.
\end{equation}

\begin{equation}
\textbf{E}^\textrm{TE}(\boldsymbol{\mathfrak{r}})=a_2(z) \, \boldsymbol{\mathcal{E}^\textrm{TE}(\boldsymbol{\mathfrak{r}_\bot})} \, e^{i(k_2 z-\omega_2 t)}+c.c.
\end{equation}

\noindent
Here $\mathfrak{r}$ is the 3D position vector and $\mathfrak{r}_\bot$ is the position vector in the cross-sectional plane transverse to the propagation direction along z, and we separate the transverse modal distributions ($\boldsymbol{\mathcal{E}^\textrm{TE}(\boldsymbol{\mathfrak{r}_\bot})}$ and $\boldsymbol{\mathcal{E}^\textrm{TM}(\boldsymbol{\mathfrak{r}_\bot})}$) from the modal amplitudes ($a_1(z)$ and $a_2(z)$). Since we are interested in the electro-optic effect we can write the material perturbation as an RF input field:

\begin{equation}
    \textbf{E}^\textrm{RF}(\boldsymbol{\mathfrak{r}}) = b(z) \, \boldsymbol{\mathcal{E}^\textrm{RF}(\boldsymbol{\mathfrak{r}_\bot})} \,e^{i(qz-\Omega t)}+c.c.
\end{equation}

\noindent
We can then assume that the dynamic refractive index grating (due to electro-optic modulation) created by this RF wave can be expressed under slowly varying wave approximation as:

\begin{equation}
    \Delta\boldsymbol{n_{\rm{EO}}}=\Delta n_{\rm{EO}} \, e^{i(qz-\Omega t)}+c.c.
\end{equation}

\noindent
where $\Delta\boldsymbol{n}\textsubscript{\textbf{EO}}$ is the full travelling-wave perturbation and $\Delta n\textsubscript{EO}$ is the amplitude of the change in the refractive index which can be evaluated using the second-order optical non-linearity (r\textsubscript{ijk})~\cite{Weis:1985aa}. We will revisit the phase matching in the next section (\S\ref{sec:Phasematching}) in detail. Presently, we will evaluate the refractive index perturbation produced by this RF field which only depends on the transverse mode profiles of the optical and RF modes. For this purpose, we calculate the change in the indicatrix as:
\begin{equation}
    \hspace{-2.5pt}
    \begin{bmatrix}
        \Delta B_{1} \\
        \Delta B_{2} \\
        \Delta B_{3} \\
        \Delta B_{4} \\
        \Delta B_{5} \\
        \Delta B_{6}
    \end{bmatrix}
    =
    \begin{bmatrix}
        r_{11} & r_{12} & r_{13} \\
        r_{21} & r_{22} & r_{23} \\
        r_{31} & r_{32} & r_{33} \\
        r_{41} & r_{42} & r_{43} \\
        r_{51} & r_{52} & r_{53} \\
        r_{61} & r_{62} & r_{63} \\
    \end{bmatrix}
    \begin{bmatrix}
        \mathcal{E}_1^\textrm{RF}(\boldsymbol{\mathfrak{r}_\bot}) \\
        \mathcal{E}_2^\textrm{RF}(\boldsymbol{\mathfrak{r}_\bot}) \\
        \mathcal{E}_3^\textrm{RF}(\boldsymbol{\mathfrak{r}_\bot}) \\
    \end{bmatrix}
    \hspace{2.5pt}
    \rm{and}
    \hspace{2.5pt}
    \begin{bmatrix}
        \boldsymbol{\mathcal{E}}^\textrm{RF}(\boldsymbol{\mathfrak{r}_\bot}) \\
    \end{bmatrix}
    =
    \begin{bmatrix}
        \mathcal{E}_1^\textrm{RF}(\boldsymbol{\mathfrak{r}_\bot}) \\
        \mathcal{E}_2^\textrm{RF}(\boldsymbol{\mathfrak{r}_\bot}) \\
        \mathcal{E}_3^\textrm{RF}(\boldsymbol{\mathfrak{r}_\bot}) \\
    \end{bmatrix}
    \label{indicatrixchange}
\end{equation}

\noindent
Where $\Delta B_i$\ is the change in the indicatrix, $\mathcal{E}_i^\textrm{Rf}(\boldsymbol{\mathfrak{r}_\bot})$ is the transverse RF field and r\textsubscript{lk} is the electro-optic tensor. In equation (\ref{indicatrixchange}), we used Neumann's principle to convert r\textsubscript{ijk} to r\textsubscript{lk} as described in Table~\ref{tab:ContractedNotation}.

		\begin{table}[ht!]
		\small
			\caption{Contracted notation by using Neumann's principle~\cite{Weis:1985aa} }
            \centering
			\begin{tabular}{ | >{\centering\arraybackslash}m{1cm} |  >{\centering\arraybackslash}m{1cm}  >{\centering\arraybackslash}m{1cm}  >{\centering\arraybackslash}m{1cm}  >{\centering\arraybackslash}m{1cm}  >{\centering\arraybackslash}m{1cm}  >{\centering\arraybackslash}m{1cm}| }
				
				\hline
	   		\textbf{ij} & 11 & 22 & 33 & 23,32 & 31,13 & 12,21\\
                \hline
                \textbf{l} & 1 & 2 & 3 & 4 & 5 & 6\\
                \hline
            \end{tabular}%
   
            \label{tab:ContractedNotation}
		\end{table}

\noindent 
We can now calculate the refractive index change by the first order perturbation theory~\cite{Weis:1985aa,Balram:14},
\begin{equation}
    \Delta n_{\rm{EO}} = -\frac{\epsilon_0 n^5}{2}
    \frac{\int
        \begin{bmatrix}
            \boldsymbol{\mathcal{E}}^\textrm{TE}(\boldsymbol{\mathfrak{r}_\bot}) 
        \end{bmatrix}^H
        \begin{bmatrix}
            \Delta B_1 & \Delta B_6 & \Delta B_5 \\
            \Delta B_6 & \Delta B_2 & \Delta B_4 \\
            \Delta B_5 & \Delta B_4 & \Delta B_3
        \end{bmatrix}
        \begin{bmatrix}
            \boldsymbol{\mathcal{E}}^\textrm{TM}(\boldsymbol{\mathfrak{r}_\bot}) 
        \end{bmatrix}
        d\boldsymbol{\mathfrak{r}_\bot}+c.c.
    }
    {
        \int
            (\boldsymbol{\mathcal{E}^\textrm{TE}(\boldsymbol{\mathfrak{r}_\bot})}+\boldsymbol{\mathcal{E}^\textrm{TM}(\boldsymbol{\mathfrak{r}_\bot})})
            \cdot
            (\boldsymbol{\mathcal{D}^\textrm{TE}(\boldsymbol{\mathfrak{r}_\bot})}+\boldsymbol{\mathcal{D}^\textrm{TM}(\boldsymbol{\mathfrak{r}_\bot})}) 
            \, 
            d\boldsymbol{\mathfrak{r}_\bot}
    }
    \label{overlapintegral}
\end{equation}
Here, n is the refractive index of the material, and H is the Hermitian operator. Also,
\begin{equation}
    \begin{bmatrix}
        \boldsymbol{\mathcal{E}}^\textrm{TE}(\boldsymbol{\mathfrak{r}_\bot}) \\
    \end{bmatrix}
        =
    \begin{bmatrix}
        \mathcal{E}_1^\textrm{TE}(\boldsymbol{\mathfrak{r}_\bot}) \\
        \mathcal{E}_2^\textrm{TE}(\boldsymbol{\mathfrak{r}_\bot}) \\
        \mathcal{E}_3^\textrm{TE}(\boldsymbol{\mathfrak{r}_\bot}) \\
    \end{bmatrix} 
    \quad \rm{and} \quad
    \begin{bmatrix}
        \boldsymbol{\mathcal{E}}^\textrm{TM}(\boldsymbol{\mathfrak{r}_\bot}) \\
    \end{bmatrix}
        =
    \begin{bmatrix}
        \mathcal{E}_1^\textrm{TM}(\boldsymbol{\mathfrak{r}_\bot}) \\
        \mathcal{E}_2^\textrm{TM}(\boldsymbol{\mathfrak{r}_\bot}) \\
        \mathcal{E}_3^\textrm{TM}(\boldsymbol{\mathfrak{r}_\bot}) \\
    \end{bmatrix}
\end{equation}

\noindent
Here we see that $\Delta n\textsubscript{EO}$ is a function of the selection rules between the photon's initial and final state (which depends on the transverse optical modal shape and the change in the indicatrix). We can understand this by using a simplified overlap integral which can be derived considering equation (\ref{overlapintegral}). For instance, the first component of the integral in (\ref{expandednotation}) is:
\begin{equation}
  \begin{gathered}[b]
    (\Delta B_1\mathcal{E}_1^\textrm{TM}(\boldsymbol{\mathfrak{r}_\bot})+\Delta B_6\mathcal{E}_2^\textrm{TM}(\boldsymbol{\mathfrak{r}_\bot})+\Delta B_5\mathcal{E}_3^\textrm{TM}(\boldsymbol{\mathfrak{r}_\bot}))\mathcal{E}_1^\textrm{TE}(\boldsymbol{\mathfrak{r}_\bot})= \\
    \mathcal{E}_1^\textrm{TM}\mathcal{E}_1^\textrm{TE}\mathcal{E}_1^\textrm{RF}r_{11}+\mathcal{E}_1^\textrm{TM}\mathcal{E}_1^\textrm{TE}\mathcal{E}_2^\textrm{RF}r_{12}+\mathcal{E}_1^\textrm{TM}\mathcal{E}_1^\textrm{TE}\mathcal{E}_3^\textrm{RF}r_{13} \\
    +\mathcal{E}_2^\textrm{TM}\mathcal{E}_1^\textrm{TE}\mathcal{E}_1^\textrm{RF}r_{61}+\mathcal{E}_2^\textrm{TM}\mathcal{E}_1^\textrm{TE}\mathcal{E}_2^\textrm{RF}r_{62}+\mathcal{E}_2^\textrm{TM}\mathcal{E}_1^\textrm{TE}\mathcal{E}_3^\textrm{RF}r_{63}\\
    +\mathcal{E}_3^\textrm{TM}\mathcal{E}_1^\textrm{TE}\mathcal{E}_1^\textrm{RF}r_{51}+\mathcal{E}_3^\textrm{TM}\mathcal{E}_1^\textrm{TE}\mathcal{E}_2^\textrm{RF}r_{52}+\mathcal{E}_3^\textrm{TM}\mathcal{E}_1^\textrm{TE}\mathcal{E}_3^\textrm{RF}r_{53}
  \end{gathered}
  \label{expandednotation}
\end{equation}

\noindent
By extension from equation (\ref{expandednotation}), we can now simplify equation (\ref{overlapintegral}) into a more compact form:
\begin{equation}
    \Delta n_{\rm{EO}} \propto \int \sum_{ijk} \mathcal{E}_i^\textrm{TM}(\boldsymbol{\mathfrak{r}_\bot}) \, \mathcal{E}_j^\textrm{TE}(\boldsymbol{\mathfrak{r}_\bot}) \,\textrm{r}_{ijk} \, \mathcal{E}_k^\textrm{RF}(\boldsymbol{\mathfrak{r}_\bot}) \, d\boldsymbol{\mathfrak{r}_\bot}
    \label{compactoverlap}
\end{equation}
as written in the main text. From this equation (\ref{compactoverlap}), we can easily see that to create coupling between two orthogonal states (such as TE and TM modes), we need non-zero r\textsubscript{ij} with $i > 4$ to produce a change in the indicatrix for the off diagonal elements in equation (\ref{overlapintegral}). Finally, we can express the electro-optic coupling rate per modal amplitude of the RF wave as,
\begin{equation}
g\textsuperscript{'}=\frac{\omega_0\Delta n_{\rm{EO}}}{n\textsubscript{g}v\textsubscript{g}}
\end{equation}

\noindent
Here, the coupling rate $g'$ is in the units of rad/m, $\omega_0$ is the optical frequency (we assume that $\omega_0 = \omega_1 \approx \omega_2 \gg \Omega$), $n_g=\sqrt{n_g^{\rm{TM}}n_g^{\rm{TE}}}$ is the geometric mean of the group indices of the modes, and $v\textsubscript{g}=\sqrt{v_g^{\rm{TM}}v_g^{\rm{TE}}}$ is the geometric mean of the group velocities.

\vspace{12pt}

\section{Phase-matching considerations and predicted spectral characteristics}
\label{sec:Phasematching}

The spatial evolution of the three-wave mixing process (i.e. the interaction between the optical fields and the RF field) is described by a well-known set of coupled mode equations~\cite{Kharel_2016,Kittlaus_2018}. 
Presently, we assume that the process is only phase-matched for anti-Stokes scattering (see \S{\ref{sec:EOrate}}) from the input optical probe (carrier) signal $a_1$ to the anti-Stokes signal $a_2$, and an RF field b inducing the interaction. For this derivation, we assume that the RF excitation amplitude is uniform throughout the waveguide, the carrier signal is not depleted, and the optical fields do not interfere with the RF electrodes, i.e. that the beat signal of the optical waves does not excite any RF wave. Then, the governing equation for the anti-Stokes scattered signal can be written as:

\begin{equation}
    \frac{\partial a_2}{\partial z}=i k_2 a_2 - i g' \, a_{1,0} \, b_0 \, e^{i(k_1+q)z}
    \label{antiStokesfinaldiff}
\end{equation}

\begin{figure}[t]
    \centering
    \includegraphics[width=0.4\textwidth]{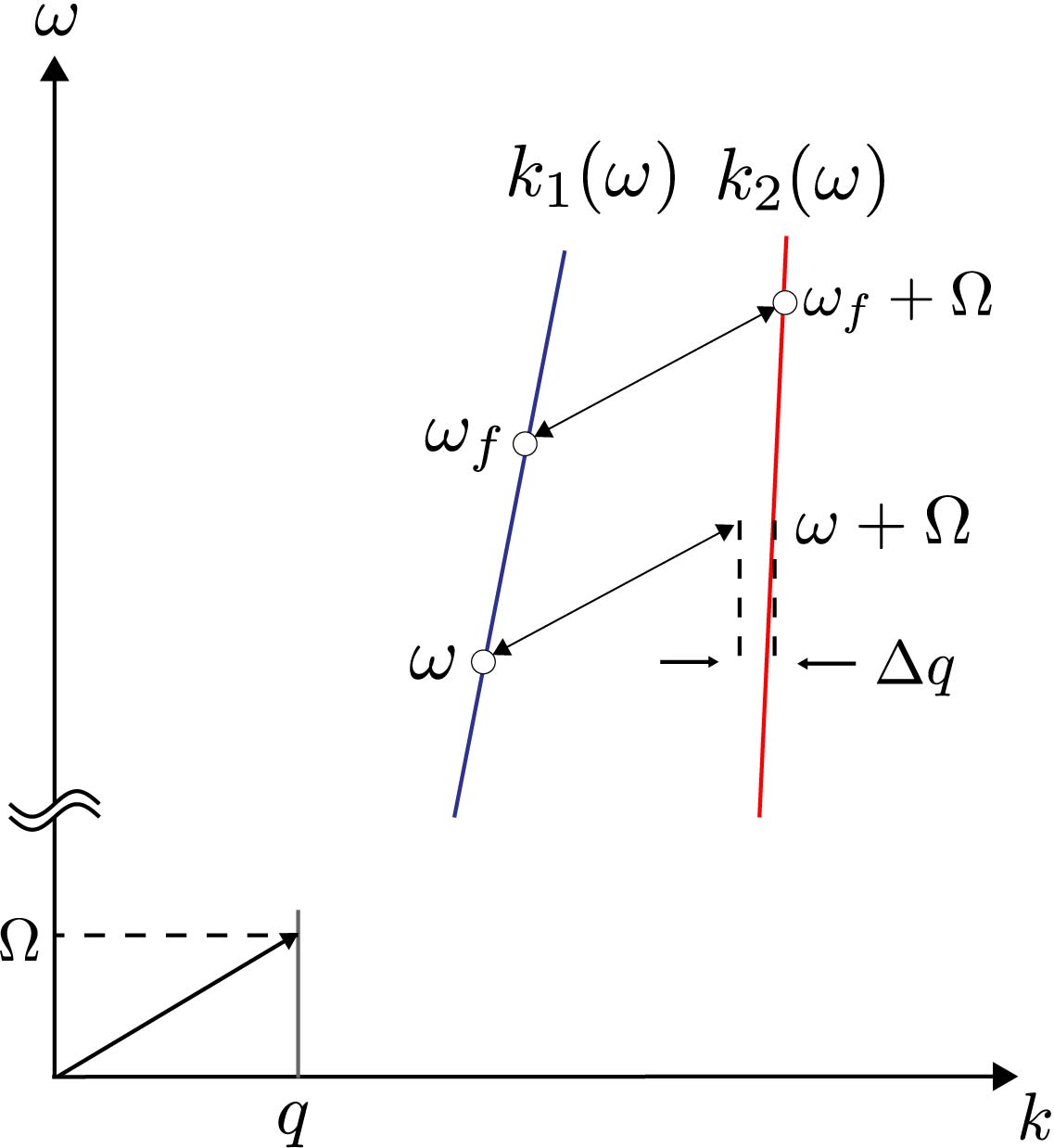}
    \caption{
    \textbf{Phase matching condition for the anti-Stokes processes:}
    The exact phase matching can be realized around $\omega_f$, and for the other optical frequencies the response follows a sinc\textsuperscript{2} function due to the group index difference.
    }
    \label{PhaseMatching}
\end{figure}

\noindent
Here, $k_1$ and $k_2$ are the wavevectors of the carrier and scattered signals, and $q$ represents the momentum of the RF excitation. Due to our assumptions, the carrier signal and the RF signal have the form $a_1(z)=a_{1,0} \, e^{ik_1z} $ and $b(z)=b_0 \, e^{iqz} $ respectively where $a_{1,0}$ and $b_0$ are the  corresponding amplitudes. 
Equation (\ref{antiStokesfinaldiff}) can be solved by assuming a solution of the form $a_2(z)  =\bar{a_2}(z) e^{ik_2z}$ where $\bar{a_2}(z)$ 
represents the spatially varying amplitude of the anti-Stokes signal, and the exponential term represents the propagation. Solving for $\bar{a_2}(z)$ we get:

\begin{equation}
    \frac{\partial \bar{a_2}}{\partial z}
    =
    \left(\frac{\partial a_2}{\partial z}-ik_2a_2\right)e^{-ik_2z}
    =
    -ig_{eo}a_{1,0}e^{i(k_1+q-k_2)z}
    \label{aftersubs}
\end{equation}
\noindent
where we define electro-optic coupling rate as $g_{eo}=g'b_0$. Then, integrating the both sides of equation~(\ref{aftersubs}), we find evolution of the anti-Stokes scattered signal at the output at position $\mathcal{L}$, i.e. $\bar{a_2}(\mathcal{L})$ as:
\begin{equation}
    \left| 
        \frac{ \bar{a_2}(\mathcal{L}) }{a_{1,0}} \right|^2=(g_{eo}\mathcal{L})^2 \rm{sinc}^2((k_1+q-k_2)\mathcal{L}/2)
    \label{sincresponse}
\end{equation}
\noindent
We see that the overall response follows a $\rm{sinc}^2$ behaviour depending on the total interaction length ($\mathcal{L}$), and the phase matching condition ($k_1+q-k_2$). The scattered power increases as the interaction length increases, and it is maximized for the perfectly phase-matched condition ($k_1+q-k_2=0$). Notably, this phase matching condition is only perfectly satisfied at a one optical frequency $\omega_f$ (see Fig.~\ref{PhaseMatching}) associated with the individual optical dispersions where the following relation is met:
\begin{equation}
q = k_{2}(\omega_f+\Omega)-k_{1}(\omega_f)
\end{equation}
\noindent
For any other probe frequency $\omega$ we can write the momentum mismatch as (see Fig.~\ref{PhaseMatching}), 
\begin{equation}
\Delta k = k_{2}(\omega+\Omega)-k_{1}(\omega)
\end{equation}

\noindent
Now the phase matching condition can be re-written to incorporate the dispersive nature of the process as:
\begin{equation}
\Delta k -q = \Delta q = (k_{2}(\omega+\Omega)-k_{2}(\omega_f+\Omega))-(k_{1}(\omega)-k_{1}(\omega_f))
\label{phasemismatchcomp}
\end{equation}
\noindent
Assuming linear dispersion, we can approximate (\ref{phasemismatchcomp}) as:

\begin{equation}
\Delta q\approx \frac{\partial k_2}{\partial\omega}(\omega-\omega_f)-\frac{\partial k_1}{\partial\omega}(\omega-\omega_f)=\frac{n_{g,2}-n_{g,1}}{c}\Delta\omega
\label{phasemismatch}
\end{equation}

\noindent
Here, $n_{g,1}$ and $n_{g,2}$ are the group velocities of the two modes and $\Delta\omega$ is the detuning from the phase matched optical frequency ($\omega_f$). As we substitute eqn.~(\ref{phasemismatch}) to equation~(\ref{sincresponse}), we can directly observe that the modulation efficiency changes for different optical frequencies. From here we can calculate the full width half maximum of $\rm{sinc}^2(x)$ function which is realized when $x\approx0.4425\pi$ (i.e. $\Delta q\mathcal{L}/2\approx0.4425\pi$ for our case). By using (\ref{phasemismatch}), we can find the optical bandwidth as:

\begin{equation}
    \Delta\omega_{BW}\approx\frac{4\cdot0.4425\cdot \pi c}{\mathcal{L}|n_{g,2}-n_{g,1}|}
    \label{bandwidth}
\end{equation}
\noindent
Here, the additional factor of 2 is introduced since we measure the total span of the scattering bandwidth. In terms of wavelength, this bandwidth is equivalent to,

\begin{equation}
    \Delta\lambda_{BW}\approx\frac{2\cdot0.4425\cdot\lambda_0^2}{\mathcal{L}|n_{g,2}-n_{g,1}|}
    \label{opticalbandwidth}
\end{equation}

\noindent
For our waveguide design, we estimated the group indices for the TE (mode 2) and TM (mode 1) modes as 2.339 and 2.536 respectively by means of FEM simulations. By using equation (\ref{opticalbandwidth}), we estimate the bandwidth for a device with length $\mathcal{L}$ = 1.08 mm as $\Delta\lambda_{BW} = 2.53$ nm which is very similar to the experimental observation of $\approx$ 2.42 nm.

\vspace{12pt}
The above mode conversion analysis is based on the non-depleted pump approximation, and it can be easily be extended for larger conversion rates with coupled mode equations by additionally considering the depletion of the carrier signal. The analytical solution for this type of three-wave mixing process is well known~\cite{Yariv_1973}, and the scattered power for our case can be expressed as:

\begin{equation}
    \left|\frac{\bar{a_2}(\mathcal{L})}{a_{1,0}}\right|^2=\frac{4g_{eo}^2}{(4g_{eo}^2+\Delta q^2)}sin^2((4g_{eo}^2+\Delta q^2)^{1/2}\mathcal{L}/2)
    \label{Largesignal}
\end{equation}

\noindent
Similarly, $a_{1,0}$ is the initial carrier signal amplitude. The only difference between (\ref{sincresponse}) and (\ref{Largesignal}) is the hybridization of the optical modes under large conversion rates. The phase matching condition for the optical waves do not change and still depend on the group index difference.

\section{Calculation of non-reciprocal bandwidth}
\label{sec:Nonreciprocal}

\begin{figure}[th!]
    \centering
    \includegraphics[width=0.8\textwidth]{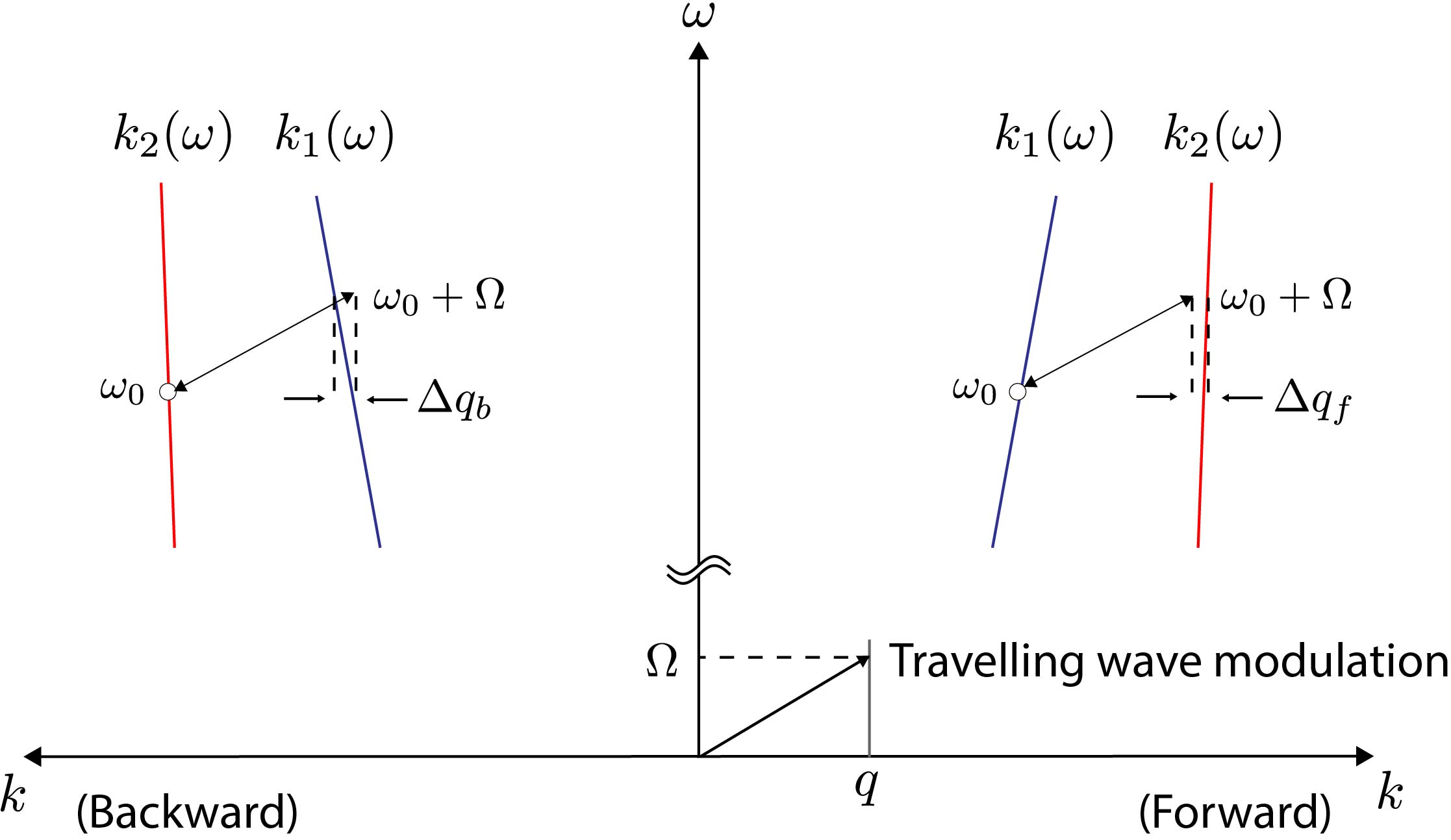}
    \caption{
    \textbf{Phase matching conditions for the forward and backward anti-Stokes processes:}
    The anti-Stokes scattering process for the forward and backward directions shows different requirements for phase matching condition.
    }
    \label{FBPmatching}
\end{figure}

\noindent
In our system, non-reciprocity arises as the applied RF field breaks the time reversal symmetry~\cite{Kittlaus_2018}. To explore this concept, we again consider the anti-Stokes processes for the forward and backward directions at some optical probing frequency $\omega_0$ ( similar equations can be derived for the Stokes process) as shown in Fig.~{\ref{FBPmatching}}. For the forward direction, for a probe injected into mode 1, the phase mismatch is given by
\begin{equation}
\Delta q_f=k_{2}(\omega_0+\Omega)-k_{1}(\omega_0)-q ~.
\end{equation}

\begin{figure}[th!]
    \centering
    \includegraphics[width=0.8\textwidth]{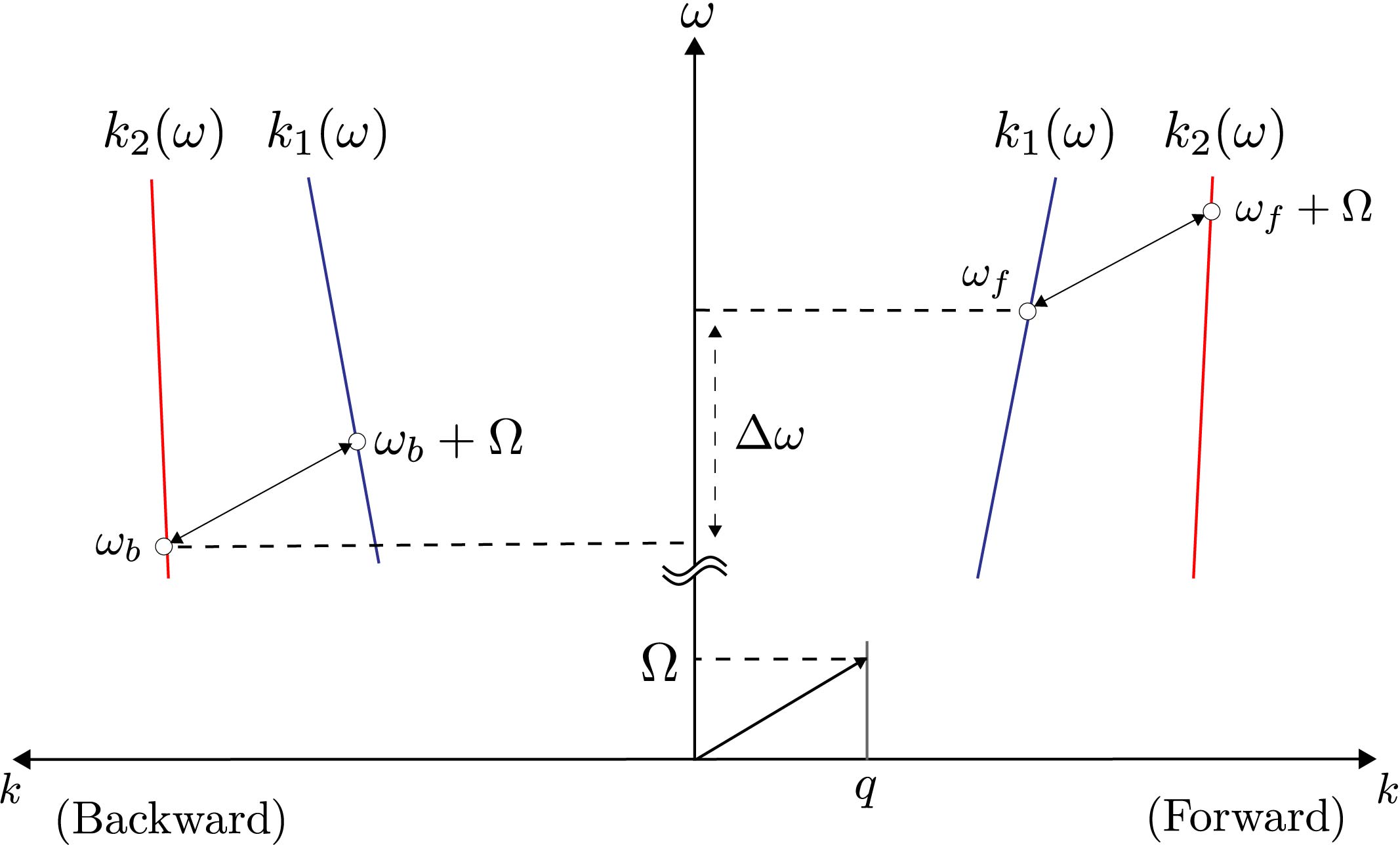}
    \caption{
    \textbf{The spectral shift for the anti-Stokes processes:}
    The group index difference between the optical modes results in different phase matching frequencies for the forward and backward directions.
    }
    \label{FBFdifference}
\end{figure}

\noindent
Similarly, in the backward direction for a probe injected to mode 2, the phase mismatch is,
\begin{equation}
\Delta q_b=k_{2}(\omega_0)-k_{1}(\omega_0+\Omega)-q
\end{equation}
Here, the probing mode in the backward direction is different since the forward directed momentum of the applied modulation never allows for anti-Stokes scattering from mode 1 to mode 2 in this direction.
\noindent
Then, taking difference between these two equations, we find that,

\begin{equation}
\Delta q_{f}-\Delta q_{b}=(k_{2}(\omega_0+\Omega)-k_{2}(\omega_0))-(k_{1}(\omega_0)-k_{1}(\omega_0+\Omega))
\end{equation}
\noindent
If we assume linear dispersion for the frequency of interest, the above equation simplifies by using Taylor's expansion,

\begin{equation}
\Delta q_{f}-\Delta q_{b}\approx\frac{\Omega}{c}(n_{g,2}+n_{g,1})
\label{frequencyshift}
\end{equation}
\noindent
From Eqn.~(\ref{frequencyshift}), we see that the phase matching conditions for the forward and backward directions depend on the modulation frequency and the group index. Importantly, this shows that the anti-Stokes process is not reciprocal (similarly the Stokes process). Since the forward and backward directions require different phase matching conditions ($\Delta q_{f}$ and $\Delta q_{b}$), the phase matching frequencies are also different from each other (Fig.~\ref{FBFdifference}).

\vspace{12pt}

\noindent
The difference between $\omega_f$ and $\omega_b$ can be found by considering the perfect phase matching condition for the anti-Stokes process in the backward direction.
When this occurs (i.e. $\Delta q_b=0$), the forward anti-Stokes process accumulates some phase ($\Delta q_f\neq0$ ) which can be predicted by using equation ({\ref{frequencyshift}}). Similarly, the same phase accumulation can be calculated by equation ({\ref{phasemismatch}}) which identifies the optical frequency $\omega_f$ where the phase matching will be perfect, i.e. $\Delta q=\Delta q_f$. Thus by using equations ({\ref{phasemismatch}}) and ({\ref{frequencyshift}}) we can calculate the frequency separation as:

\begin{equation}
(\omega_{b}-\omega_{f})\frac{n_{g,1}+n_{g,2}}{c}=\Omega\frac{n_{g,2}-n_{g,1}}{c}
\end{equation}

\begin{equation}
\omega_{b}-\omega_{f}=\frac{n_{g,1}+n_{g,2}}{n_{g,2}-n_{g,1}}\Omega
\label{spectralshift}
\end{equation}

\noindent
For significant non-reciprocity, this spectral shift ({\ref{spectralshift}}) should be larger than modulation bandwidth ({\ref{bandwidth}}), for which we require,

\begin{equation}
\frac{n_{g,1}+n_{g,2}}{|n_{g,2}-n_{g,1}|}\Omega\gg\frac{4\cdot0.4425\cdot \pi c}{\mathcal{L}|n_{g,2}-n_{g,1}|}
\end{equation}

\noindent
which is equivalent to

\begin{equation}
\mathcal{L}\gg\frac{4\cdot0.4425\cdot \pi c}{(n_{g,1}+n_{g,2})\Omega}
\label{overalllength}
\end{equation}

\noindent
As can be seen from eqn.(\ref{overalllength}), the nonreciprocal contrast enhances with larger values of $\mathcal{L}$ or larger values of $\Omega$. In practice, we limit $\mathcal{L}$ to stay at or below the length for full conversion (guided by equation {\ref{Largesignal}}). For our waveguide design, we estimated the group indices for the TE (mode 2) and TM (mode 1) modes as 2.339 and 2.536 respectively. By using our experiment parameters, we estimate the spectral shift of a device with length of $\mathcal{L}=1.08$ mm and a modulation frequency of $\Omega=6$ GHz to be to be 0.302 nm which is close to the experimental observation of $\approx$ 0.3 nm (See Fig.~\ref{fig:5})

\section{Generation of a traveling wave modulation via spatial sampling}
\label{sec:Syntheticmomentum}

\begin{figure}[th!]
    \centering
    \includegraphics[width=\textwidth]{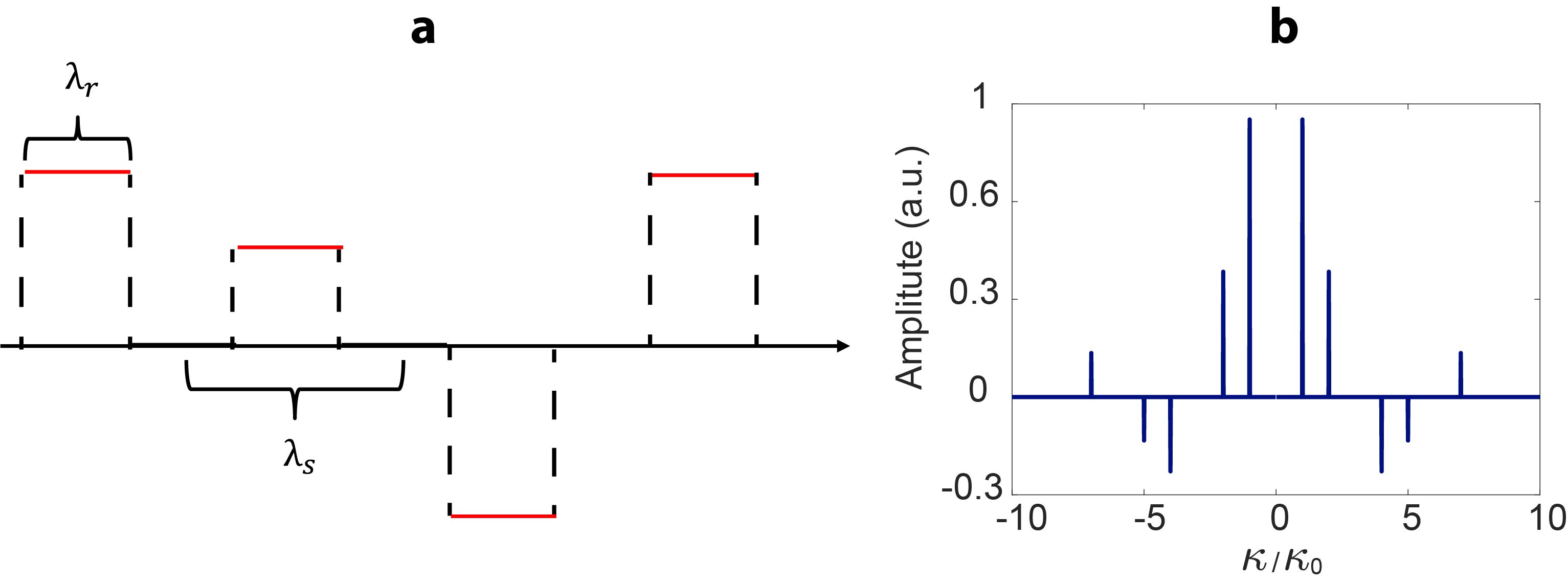}
    \centering
    \caption{
    \textbf{Description of the synthetically generated travelling wave}
    \textbf{(a)} Three point sampling of a sinusoidal wave with a $\rm{rect}(x)$ function.
    \textbf{(b)} The resulting signal spectrum after three point sampling of a sinusoidal wave with $\rm{rect}(x)$ function 
    }
    \label{Syntheticwave}
\end{figure}

In our experimental realization, we used a split electrode structure to realize the required traveling wave modulation. Each electrode (of pitch $L$) is provided a single tone RF stimulus at frequency $\Omega$ with a relative phase shift of $\Delta\phi$ between electrodes. The incremental phase of this array ($\Delta\phi$) and the electrode length ($L$) determines the momentum of our synthetic wave~\cite{Sounas:14} ($q=\Delta\phi/L$) bridging the momentum gap for the inter-polarization scattering process. In this section, we analyze the momentum spectrum of our phased array. For this purpose, we consider a continuous periodic spatial signal of $g(x)$ sampled by a train of $\rm{rect}(x)$~\cite{Oppenheim_signal} functions representing our split electrode structure. We can write this ``rectangular sampled'' signal (illustration in Fig.~{\ref{Syntheticwave}a}) as follows:

\begin{equation}
g_{s}(x)=\sum_{n}^{} g(n\lambda_{s}) \, \rm{rect}\left(\frac{x-n\lambda_{s}}{\lambda_{r}}\right)
\label{sampledequation}
\end{equation}
\noindent
Here, $\lambda_{r}$ represents the width of the $\rm{rect}(x)$ function (i.e. single electrode length), and $\lambda_{s}$ represents the period for the sampling which is equal to the period of a single electrode in our structure (See Fig.~\ref{Syntheticwave}a). Equation (\ref{sampledequation}) can be also written as:

\begin{equation}
g_{s}(x)=\rm{rect}\left(\frac{x}{\lambda_{r}}\right)\circledast\sum_{n}^{} g(n\lambda_{s})\delta(x-n\lambda_{s})
\label{sinusoidsampled}
\end{equation}
\noindent
Here, `$\circledast$' represents convolution operation. Taking the Fourier transform \cite{Oppenheim_signal} of the equation~(\ref{sinusoidsampled}):

\begin{equation}
G_{s}(\kappa)=\frac{sin(\pi\kappa/\kappa_{r})}{\pi\kappa}\frac{1}{\lambda_{s}}\sum_{n}^{} G(\kappa-n\kappa_{s})
\label{finalsignal}
\end{equation}

\noindent
Here, we see that the final signal is of a shifted spectral response of the original signal which is modified by the $\rm{sinc}(x)$ envelope. To explore spectral harmonics of this sampled wave, we plot the equation (\ref{finalsignal}) for a sinusodial $g(x)$ signal 
with a period of $3\lambda_s$ in Fig.~\ref{Syntheticwave}b which mimics our phased array. We see that the power is mainly reserved in the first-order lobes and the higher order sidebands shows a decaying response due to the $\rm{sinc}$ function. In our device design, we use these main sidebands to supply the required momentum for efficient TE-TM scattering process. 

\section{Electrical analysis of the electrodes and cut-off frequency}
\label{sec:Electricalmodel}

\begin{figure}[th!]
    \centering
    \includegraphics[width=0.9\textwidth]{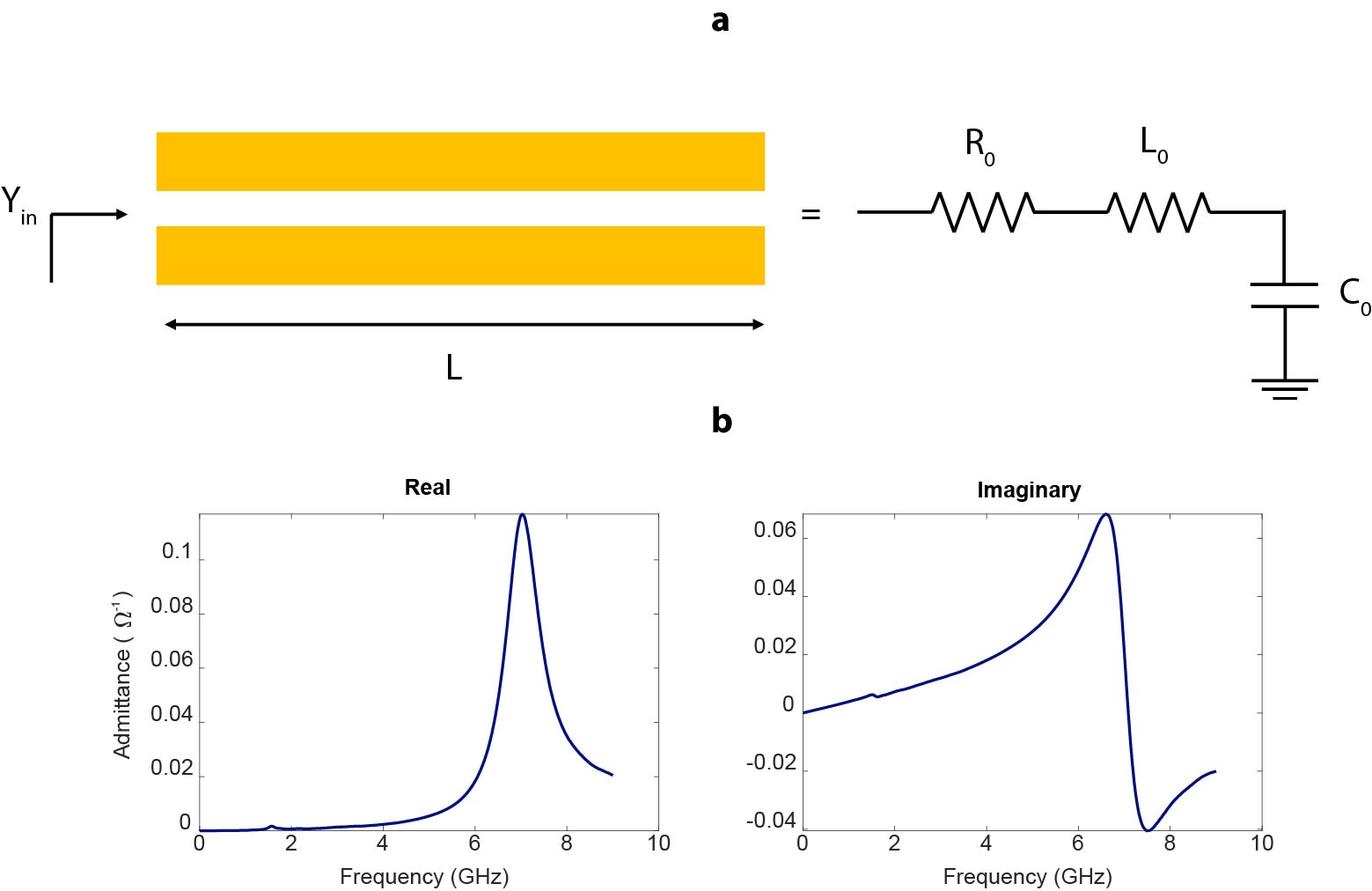}
    \caption{
    \textbf{Electrical modelling of the RF electrodes}
    \textbf{(a)} Circuit equivalent of the electrodes for standing wave region~\cite{Kim_Lee_Yun_2012}.
    \textbf{(b)} Admittance measurements of the electrodes by a vector network analyzer.
    }
    \label{Impedancemeasurements}
\end{figure}

\noindent
For our split electrode design, we must ensure that the RF excitation is in the standing wave regime as it provides the flexibility to engineer the synthetic momentum with lithographically defined features (electrode pitch and sampling).
In contrast, the propagating wave regime {\cite{Ghione_2009}} is useful for phase modulators to satisfy the momentum matching at very high frequencies {\cite{Ghione_2009,Wang_2018,Zhu:21}} (above 100 GHz).
In the standing wave regime the electrode will act as a simple RLC circuit~\cite{Kim_Lee_Yun_2012} (See Fig.~\ref{Impedancemeasurements}a). 
It is important for us to understand the cut-off frequency over which the standing wave behavior will be maintained. 
Beyond this frequency, the transferred voltage to the modulating capacitor reduces, leading to inefficient electro-optic modulation. 

\begin{figure}[h!t]
    \centering
    \includegraphics[width=0.9\textwidth]{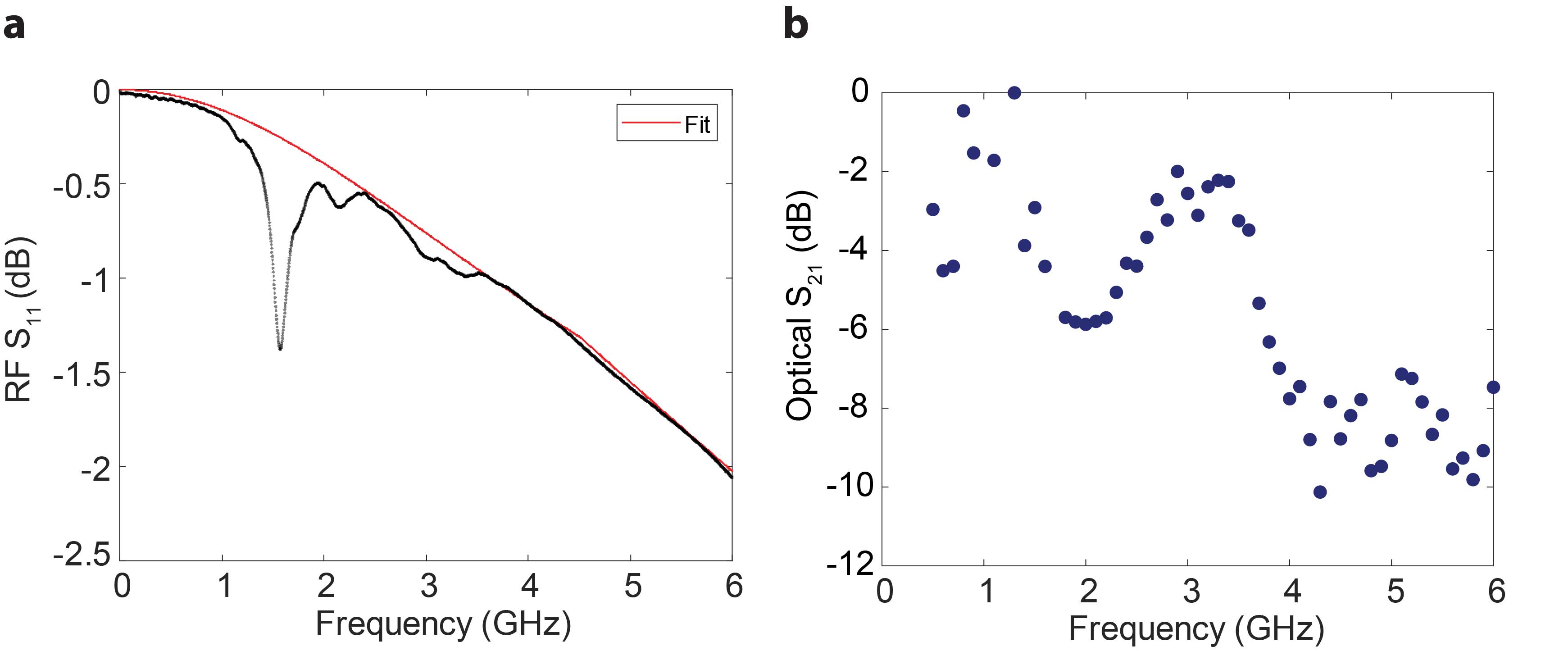}
    \caption{
    \textbf{Modulation response}
    \textbf{(a)} The reflection coefficient (S\textsubscript{11}) of the electrodes is measured by a vector network analyzer demonstrating an RC response ensuring standing wave approximation. 
    \textbf{(b)} Optical S\textsubscript{21} parameter for different optical frequencies
    }
    \label{Electricalparameters}
\end{figure}

We first measure the Y parameters (admittance parameters) for our electrodes. As the inductive component is not significant at low frequencies, we can estimate the resistivity and capacitance of our structure with measurements at 10 kHz. Here, we estimation $C\textsubscript{0}=0.6$ $pF$ and $R\textsubscript{0}= 9$ $\Omega$. The capacitance value agrees well with the finite element simulations, however the resistance is larger than the expected value of $3$ $\Omega$. This is likely related to the quality of the gold film used in the electrodes since it is deposited by e-beam evaporation. We then estimated the inductance value as $L\textsubscript{0}=0.83$ $nH$ by using the standing wave resonance of the structure, as can be seen in the measurements shown in Fig.~\ref{Impedancemeasurements}b. With these parameters, we can confirm that the cut-off frequency for this transmission line is relatively large ($>$10 GHz) and that at our operational frequency of 0-6 GHz the standing wave approximation is valid.
We additionally present the RF S\textsubscript{11} measurement in Fig.~\ref{Electricalparameters}a, which displays an RC response over the operational frequency range as expected. The little dip around 1.5 GHz frequency is due to a resonance in the cables caused due to back-reflection from the electrode. After measuring the electrical parameters, we additionally performed an optical S parameter analysis to understand the modulation response for different frequencies. The results of this measurement is also given in Fig.~\ref{Electricalparameters}b.

\FloatBarrier

\section{Important cross-sectional dimensions}
\label{sec:Dimensions}
\begin{figure}[th!]
    \centering
    \includegraphics[width=0.8\textwidth]{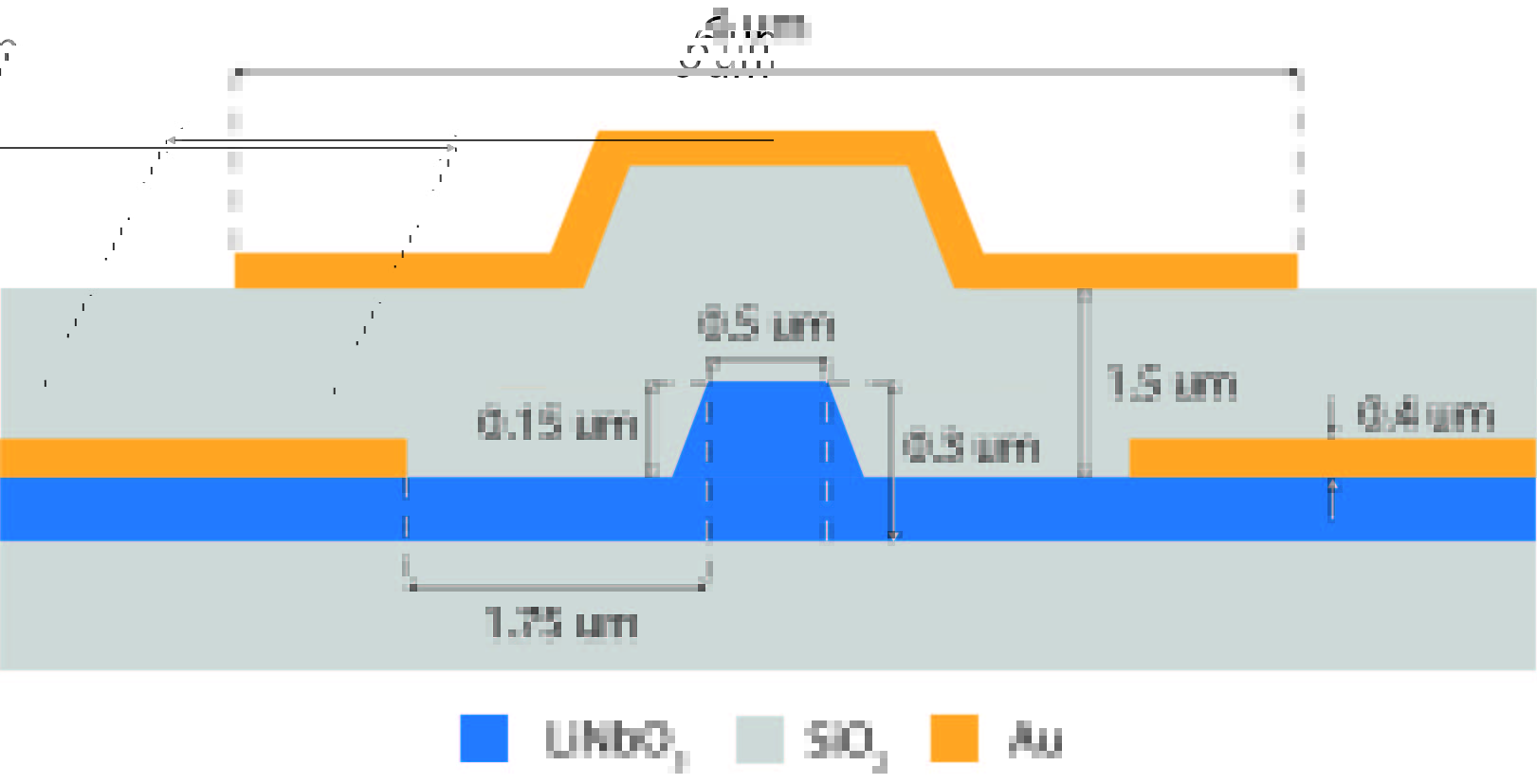}
    \caption{
    \textbf{Physical dimensions of the EO-NPR waveguide cross-section:}
    The dimension for the structure are chosen to maximize electro-optic coupling rate while minimizing the optical loss rate.
    }
    \label{Dimensions}
\end{figure}

\noindent
For our experimental demonstration, we optimized the electrode distance in order to maximize the electro-optic coupling rate while minimizing the optical loss. For our design, the bottom electrode distance to our optical structure is chosen to be 1.75 um for a waveguide width of 0.5 um. The cladding thickness (and hence the top electrode distance) is chosen to be 1.5 um since the optical fields are well confined in the vertical direction. All other dimensions of our structure are presented in Fig.\ref{Dimensions}. 

\section{Optical characterization}
\label{sec:OpticalCharacterization}

\begin{figure}[th!]
    \begin{adjustwidth}{0in}{0in}
    \centering
    \includegraphics[width=0.9\textwidth]{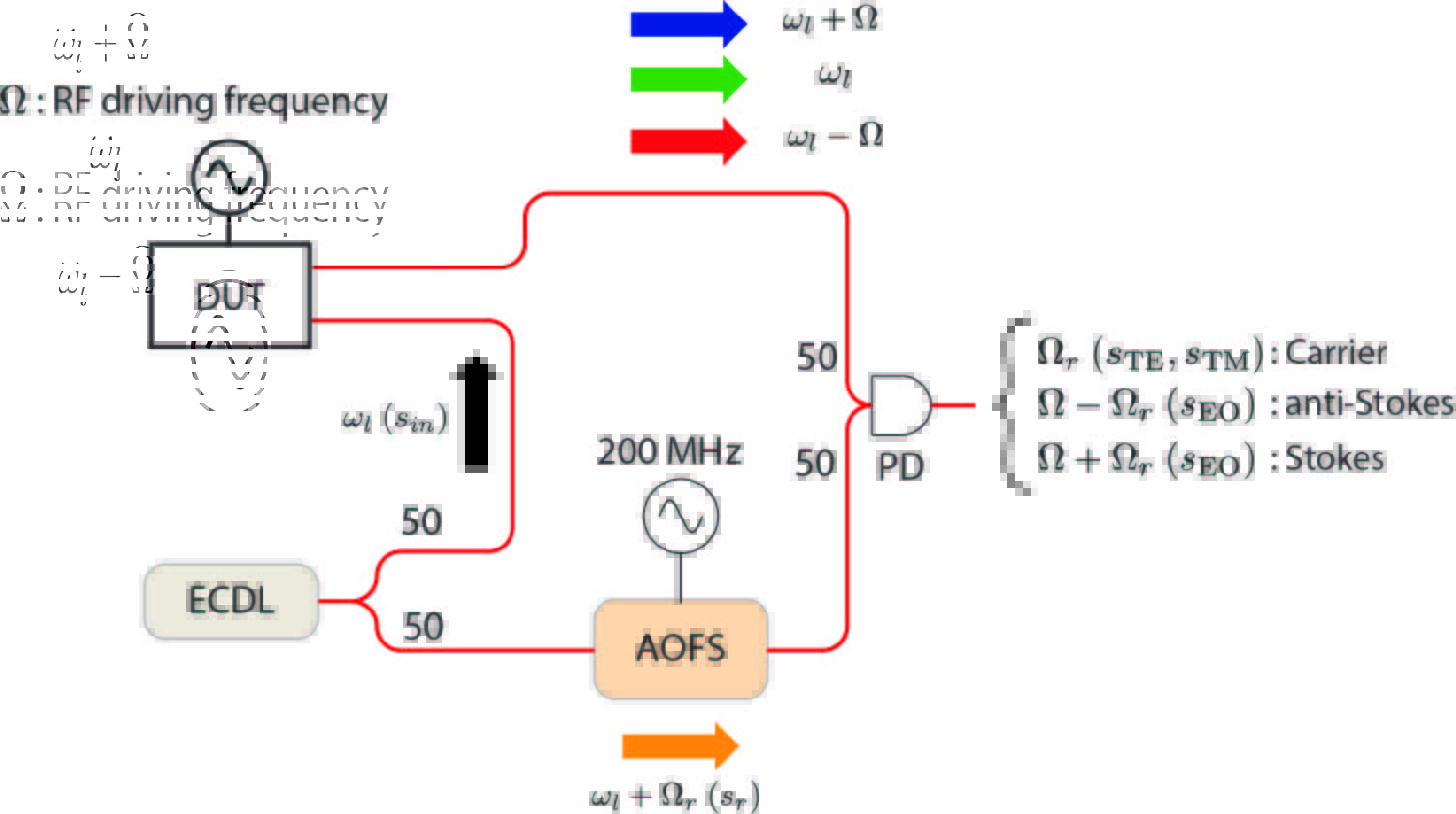}
    \end{adjustwidth}
    \caption{
    \textbf{Heterodyne detection system} ECDL: External cavity diode laser, AOFS: Acousto-optic frequency shifter, PD: Photodetector, DUT: Device under test, $\Omega_r$: Operating frequency of the AOFS (i.e. 200 MHz)
    }
    \label{finalsupp}
\end{figure}

In order to measure the Stokes, anti-Stokes, and the carrier transmission, we utilize an optical heterodyne detection system by using an acousto-optic frequency shifter (Fig.~\ref{finalsupp})

\vspace{12pt}

\begin{figure}[th!]
    \begin{adjustwidth}{-1in}{-1in}
    \centering
    \includegraphics[width=1.15\textwidth]
    {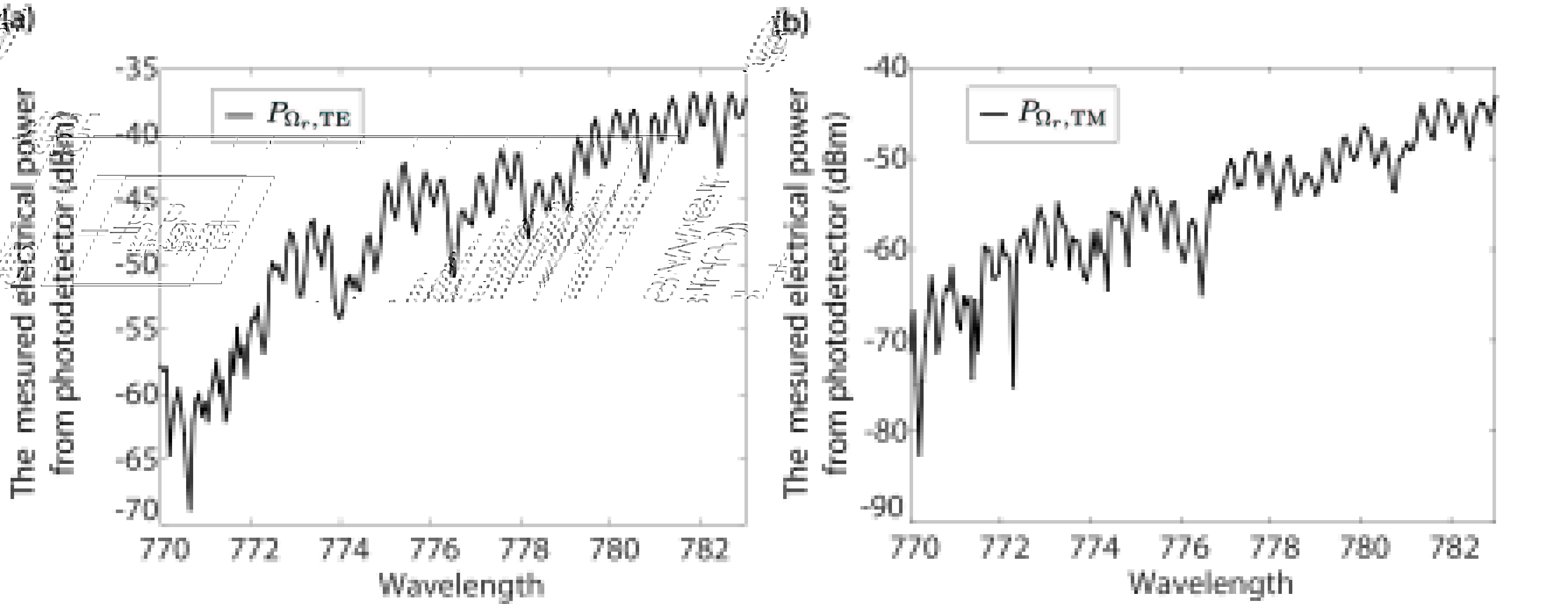}
     \end{adjustwidth}
    \caption{
    \textbf{Optical characterization}
    \textbf{(a)} The measured electrical signal for TE to TE transmission for the wavelength of interest (770 nm - 783 nm).
    \textbf{(b)} The measured electrical signal for TM to TM transmission for the wavelength of interest (770 nm - 783 nm).
    }
    \label{Electricalcarrier}
   
\end{figure}

For our experiment as given in the Main Manuscript Fig.~\ref{fig:3}, we first measure TE to TE and TM to TM transmissions through the device and gratings without any RF stimulus applied. For the TE case, we can write the optical field arriving at the photo-detector as;
\begin{equation}
s_\textrm{TE}=\frac{s_{in}e^{i\omega_lt}\sqrt{G_\textrm{TE}}\sqrt{G_\textrm{TE}}}{\sqrt{2}}+s_{r}e^{i(\omega_l+\Omega_r)t}
\end{equation}
Here, the first term represents the TE to TE transmission, and the second term represents the modulated optical signal due to AOFS. $G_\textrm{TE}$ is the power transmission coefficient for the TE gratings, and $\sqrt{2}$ appears due to our Y-splitter near the output. Similarly, for TM case, the the optical field becomes;
\begin{equation}
s_\textrm{TM}=\frac{s_{in}e^{i\omega_lt}\sqrt{G_\textrm{TM}}\sqrt{G_\textrm{TM}}}{\sqrt{2}}+s_{r}e^{i(\omega_l+\Omega_r)t}
\end{equation}
Similarly, $G_\textrm{TM}$ is the power transmission of the gratings. The resulting RF signal from the photodetector is the beat note at $\Omega_r$ and can be expressed for these two cases as;
\begin{equation}
P_{\Omega_r,\textrm{TE}}=g_{pd}|s_{r}|^2\frac{|s_{in}|^2G_\textrm{TE}^2}{2}
\label{TM_carrier}
\end{equation}
\begin{equation}
P_{\Omega_r,\textrm{TM}}=g_{pd}|s_{r}|^2\frac{|s_{in}|^2G_\textrm{TM}^2}{2}
\label{TE_carrier}
\end{equation}
Here, $g_{pd}$ is the sensitivity of the photo-detector. We plot these RF power measurements in Fig.~\ref{Electricalcarrier}.

We then perform the inter-polarization scattering experiment (Main Manuscript Fig.~\ref{fig:4}-\ref{fig:5} ) where we inject TM mode and receive TE sidebands. For this case, the measured optical spectrum at the photo-detector can be expressed as;
\begin{align}
    s_{\textrm{EO}} =& ~ g_{eo} \, \mathcal{L} \, \textrm{sinc}(\Delta q_{S}\mathcal{L}/2)\frac{s_{in} \, e^{i(\omega_l-\Omega)t} \, \sqrt{G_\textrm{TE}} \, \sqrt{G_\textrm{TM}}}{\sqrt{2}} \notag
    \\ 
    & +g_{eo} \, \mathcal{L} \, \textrm{sinc}(\Delta q_{AS}\mathcal{L}/2)\frac{s_{in}e^{i(\omega_l+\Omega)t} \, \sqrt{G_\textrm{TE}} \, \sqrt{G_\textrm{TM}}}{\sqrt{2}}
        +s_{r}e^{i(\omega_l+\Omega_r)t}
\end{align}
Here, $\Delta q_{S}$ and $\Delta q_{AS}$ are the phase mismatch for the Stokes and anti-Stokes processes. The resulting RF electrical signals from the photodetector are the beat notes of the optical reference signal with the sidebands which occur at $\Omega+\Omega_r$ and $\Omega-\Omega_r$. We can express these signals as;
\begin{equation}
P_{\Omega+\Omega_r}=g_{pd}|s_{r}|^2\left(g_{eo} \, \mathcal{L} \, \textrm{sinc}(\Delta q_{S}\mathcal{L}/2)\right)^2\frac{|s_{in}|^2 \, G_\textrm{TE} \, G_\textrm{TM}}{2}
\label{S_detector}
\end{equation}
\begin{equation}
P_{\Omega-\Omega_r}=g_{pd}|s_{r}|^2(g_{eo} \, \mathcal{L} \, \textrm{sinc}(\Delta q_{AS}\mathcal{L}/2)^2\frac{|s_{in}|^2 \, G_\textrm{TE} \, G_\textrm{TM}}{2}
\label{AS_detector}
\end{equation}
Our goal is to determine the sideband power relative to the carrier, without considering the grating efficiencies or the Y-splitter.
This can be accomplished using Eqns. (\ref{TM_carrier}) and (\ref{TE_carrier}) by taking the ratio:
\begin{equation}
\bar{P}_{\Omega+\Omega_r}=\frac{P_{\Omega+\Omega_r}}{\sqrt{P_{\Omega_r,\textrm{TE}} \, P_{\Omega_r,\textrm{TM}}}}=(g_{eo} \, \mathcal{L} \, \textrm{sinc}(\Delta q_{S}\mathcal{L}/2))^2
\label{fitting_equation_S}
\end{equation}
\begin{equation}
\bar{P}_{\Omega-\Omega_r}=\frac{P_{\Omega-\Omega_r}}{\sqrt{P_{\Omega_r,\textrm{TE}} \, P_{\Omega_r,\textrm{TM}}}}=(g_{eo} \, \mathcal{L} \, \textrm{sinc}(\Delta q_{AS}\mathcal{L}/2))^2
\label{fitting_equation}
\end{equation}

\begin{figure}[th!]
    \begin{adjustwidth}{-1in}{-1in}
    \centering
    \includegraphics[width=1.15\textwidth]{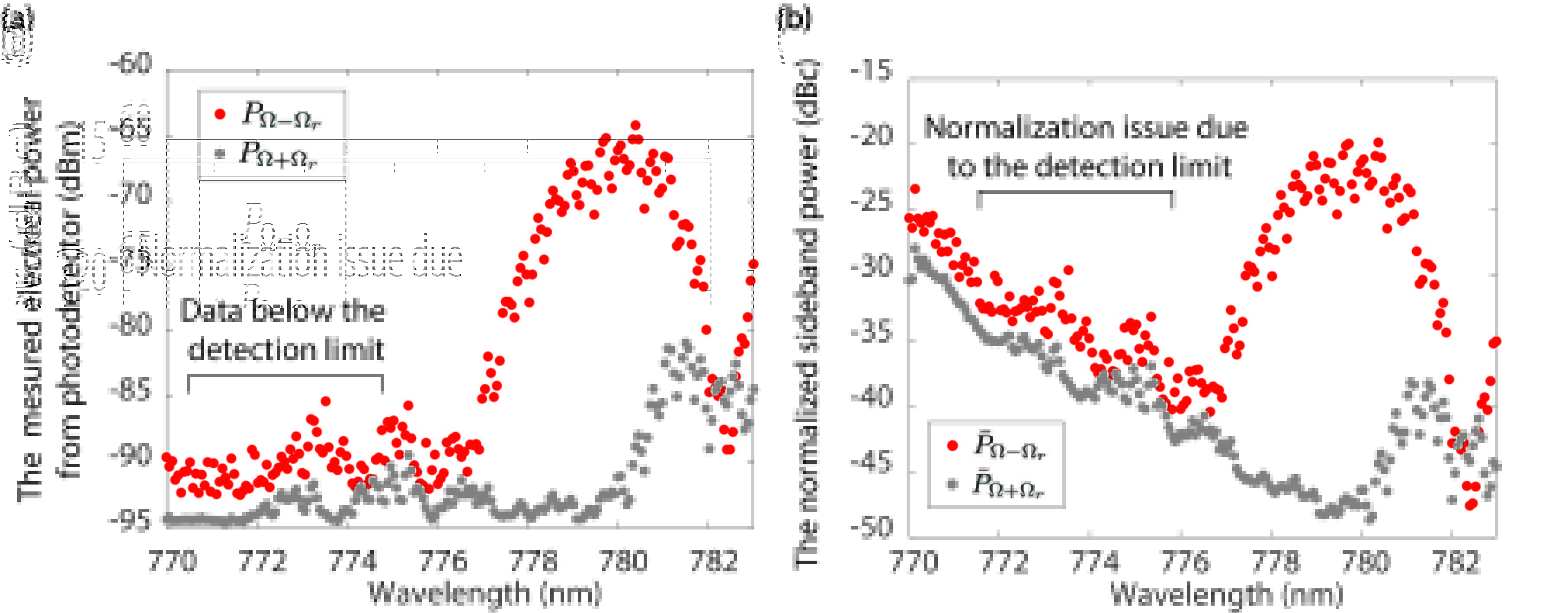}
     \end{adjustwidth}
    \caption{
    \textbf{Sideband measurements}
    \textbf{(a)} The measured electrical signal for TM to TE conversion process for the wavelength of interest (770 nm - 783 nm). 
    \textbf{(b)} The normalized electrical signal for (by using equations \ref{fitting_equation_S} and \ref{fitting_equation}) TM to TE conversion process for the wavelength of interest (770 nm - 783 nm).
    }
    \label{Electricalsideband}
   
\end{figure}

\noindent
On the dB scale, the equations (\ref{fitting_equation_S}-\ref{fitting_equation}) become;
\begin{equation}
\bar{P}_{\Omega+\Omega_r}^\textrm{dB}=P_{\Omega+\Omega_r}^\textrm{dB}-\frac{1}{2}(P_{\Omega_r,\textrm{TE}}^\textrm{dB}+P_{\Omega_r,\textrm{TM}}^\textrm{dB})
\end{equation}
\begin{equation}
\bar{P}_{\Omega-\Omega_r}^\textrm{dB}=P_{\Omega-\Omega_r}^\textrm{dB}-\frac{1}{2}(P_{\Omega_r,\textrm{TE}}^\textrm{dB}+P_{\Omega_r,\textrm{TM}}^\textrm{dB})
\end{equation}

In Fig.~\ref{Electricalsideband} we plot both the measured and the normalized electrical signals for TM to TE conversion. We see that due to the small transmission of the gratings near 770 nm, the spectrum analyzer measures noise rather than the converted signal, which causes problem during normalization. For this reason, the only reliable data is from 775 nm to 783 nm (as plotted in the main manuscript) and can be fitted with the theory. From this, we can extract the peak electro-optic modulation strength by using equations (\ref{sincresponse}, \ref{fitting_equation})  and can predict performance of a much longer device via equation (\ref{Largesignal}). 

\vspace{12pt}
By means of the experimental measurements in Fig.~\ref{Electricalsideband}, we calculate the electro-optic conversion rate to be $g_{eo}=94.21$ rad/m for our demonstrated EO-NPR with 21 dBm drive power. 
Through equation (\ref{Largesignal}) we can now estimate that the length required for full conversion will be 16.74 mm if the same $g_{eo}$ is maintained along the full length. Since the applied power is 21 dBm (for each electrode) we estimate, with the support of the electrode characterization measurements, that the voltage on the modulating electrodes is $\approx 7.1$ V allowing us to estimate the voltage-length product as $7.1 \, \rm{V} \times 1.67 \, \rm{cm}\approx 11.85 \, \rm{V}\cdot \rm{cm}$. When we normalize the conversion rate with our applied voltage, we find $\approx$ 0.13 rad/$\rm{V}\cdot\rm{cm}$ which quantifies electro-optic polarization rotation in a manner similar to the Verdet constant in magneto-optics.

\vspace{12pt}
In order to compare our EO-NPR performance with MOFE-based counterparts we use a well-established figure of merit (FoM) -- the polarization rotation rate divided by propagation loss (i.e. rad/dB) -- which is commonly used for non-reciprocal polarization rotators. Previous results from some best-in-class MOFE devices in \cite{Zhang:19} estimate a maximum FoM of 0.66 rad/dB (52.36 rad/cm with a propagation loss of 78.95 dB/cm). Similarly, for the MO devices in~\cite{Yan:20} this FoM is estimated at 0.75 rad/dB (102.97 rad/cm with a propagation loss of 137.20 dB/cm).
With our current results in lithium niobate, we achieved a comparatively much lower polarization rotation rate of 0.94 rad/cm which could be improved with larger applied RF power levels. Even so, lithium niobate offers incredibly low loss with typical loss rate $\approx 0.1$ dB/cm (our best loss rate was previously measured around 0.02 dB/cm near 780 nm~\cite{Sohn_Orsel_Bahl_2021}). Using these numbers, we can estimate a non-reciprocal polarization rotation FoM ranging between 10-50 rad/dB which is 1-2 orders of magnitude greater than on-chip MO solutions.

\FloatBarrier   %

\vspace{1cm}

{\footnotesize \putbib}
\end{bibunit}

\end{document}